%% file: main.tex
\documentclass[twocolumn]{aastex631}

\usepackage{pgffor}
\newcommand{\setnumimages}[1]{\def\numimages{#1}}
\setnumimages{24}

\begin{document}
\newcommand{\h}{$H_{Ff160W}$}
\newcommand{\logm}{log(M$_*$)}
\newcommand{\msun}{M$_\odot$}
\newcommand{\sigone}{$\Sigma_1$}
\newcommand{\bigspace}{\vspace{10mm}}


\title{JADES Ultra-red Flattened Objects: Morphologies and Spatial Gradients in Color and Stellar Populations}

\author[0000-0003-1903-9813]{Justus L.\ Gibson}
\affiliation{Department for Astrophysical and Planetary Science, University of Colorado, Boulder, CO 80309, USA}

\author[0000-0002-7524-374X]{Erica Nelson}
\affiliation{Department for Astrophysical and Planetary Science, University of Colorado, Boulder, CO 80309, USA}

\author[0000-0003-2919-7495]{Christina C.\ Williams} \affiliation{NSF’s National Optical-Infrared Astronomy Research Laboratory, 950 North Cherry Avenue, Tucson, AZ 85719 USA}

\author[0000-0002-0108-4176]{Sedona H. Price}
\affiliation{Department of Physics and Astronomy and PITT PACC, University of Pittsburgh, Pittsburgh, PA 15260, USA}

\author[0000-0001-7160-3632]{Katherine E. Whitaker}
\affil{Department of Astronomy, University of Massachusetts, Amherst, MA 01003, USA}
\affiliation{Cosmic Dawn Center (DAWN), Denmark}

\author[0000-0002-1714-1905]{Katherine A. Suess}
\thanks{NHFP Hubble Fellow}
\affiliation{Kavli Institute for Particle Astrophysics and Cosmology and Department of Physics, Stanford University, Stanford, CA 94305, USA}

\author[0000-0002-2380-9801]{Anna de Graaff}
\affiliation{Max-Planck-Institut f\"ur Astronomie, K\"onigstuhl 17, D-69117, Heidelberg, Germany}

\author[0000-0002-9280-7594]{Benjamin D.\ Johnson}
\affiliation{Center for Astrophysics $|$ Harvard \& Smithsonian, 60 Garden St., Cambridge MA 02138 USA}

\author[0000-0002-8651-9879]{Andrew J.\ Bunker }
\affiliation{Department of Physics, University of Oxford, Denys Wilkinson Building, Keble Road, Oxford OX1 3RH, UK}

\author[0000-0003-0215-1104]{William M.\ Baker}
\affiliation{Kavli Institute for Cosmology, University of Cambridge, Madingley Road, Cambridge, CB3 0HA, UK}
\affiliation{Cavendish Laboratory, University of Cambridge, 19 JJ Thomson Avenue, Cambridge, CB3 0HE, UK}

\author[0000-0003-0883-2226]{Rachana Bhatawdekar}
\affiliation{European Space Agency (ESA), European Space Astronomy Centre (ESAC), Camino Bajo del Castillo s/n, 28692 Villanueva de la Cañada, Madrid, Spain}

\author[0000-0003-4109-304X]{Kristan Boyett}
\affiliation{School of Physics, University of Melbourne, Parkville 3010, VIC, Australia}
\affiliation{ARC Centre of Excellence for All Sky Astrophysics in 3 Dimensions (ASTRO 3D), Australia}

\author[0000-0003-3458-2275]{Stephane Charlot}
\affiliation{Sorbonne Universit\'e, CNRS, UMR 7095, Institut d'Astrophysique de Paris, 98 bis bd Arago, 75014 Paris, France}

\author[0000-0002-9551-0534]{Emma Curtis-Lake}
\affiliation{Centre for Astrophysics Research, Department of Physics, Astronomy and Mathematics, University of Hertfordshire, Hatfield AL10 9AB, UK}

\author[0000-0002-2929-3121]{Daniel J.\ Eisenstein}
\affiliation{Center for Astrophysics $|$ Harvard \& Smithsonian, 60 Garden St., Cambridge MA 02138 USA}

\author[0000-0003-4565-8239]{Kevin Hainline}
\affiliation{Steward Observatory, University of Arizona, 933 N. Cherry Avenue, Tucson, AZ 85721, USA}

\author[0000-0002-8543-761X]{Ryan Hausen}
\affiliation{Department of Physics and Astronomy, The Johns Hopkins University, 3400 N. Charles St., Baltimore, MD 21218}

\author[0000-0002-4985-3819]{Roberto Maiolino}
\affiliation{Kavli Institute for Cosmology, University of Cambridge, Madingley Road, Cambridge, CB3 0HA, UK}
\affiliation{Cavendish Laboratory, University of Cambridge, 19 JJ Thomson Avenue, Cambridge, CB3 0HE, UK}
\affiliation{Department of Physics and Astronomy, University College London, Gower Street, London WC1E 6BT, UK}

\author[0000-0003-2303-6519]{George Rieke}
\affiliation{Steward Observatory and Dept of Planetary Sciences, University of Arizona 933 N. Cherry Avenue Tucson AZ 85721, USA}

\author[0000-0002-7893-6170]{Marcia Rieke}
\affiliation{Steward Observatory, University of Arizona, 933 N. Cherry Avenue, Tucson, AZ 85721, USA}

\author[0000-0002-4271-0364]{Brant Robertson}
\affiliation{Department of Astronomy and Astrophysics University of California, Santa Cruz, 1156 High Street, Santa Cruz CA 96054, USA} 

\author[0000-0002-8224-4505]{Sandro Tacchella}
\affiliation{Kavli Institute for Cosmology, University of Cambridge, Madingley Road, Cambridge, CB3 0HA, UK}
\affiliation{Cavendish Laboratory, University of Cambridge, 19 JJ Thomson Avenue, Cambridge, CB3 0HE, UK}

\author[0000-0002-4201-7367]{Chris Willott}
\affiliation{NRC Herzberg, 5071 West Saanich Rd, Victoria, BC V9E 2E7, Canada}

\begin{abstract}

One of the more surprising findings after the first year of JWST observations is the large number of spatially extended galaxies (ultra-red flattened objects, or UFOs) among the optically-faint galaxy population otherwise thought to be compact. Leveraging the depth and survey area of the JADES survey, we extend observations of the optically-faint galaxy population to an additional 112 objects, 56 of which are well-resolved in F444W with effective sizes, $R_e > 0.25”$, more than tripling previous UFO counts. These galaxies have redshifts around $2 < z < 4$, high stellar masses ($\mathrm{log(M_*/M_{\odot})} \sim 10-11$), and star-formation rates around $\sim 100-1000 \mathrm{M_{\odot}/yr}$. Surprisingly, UFOs are red across their entire extents which spatially resolved analysis of their stellar populations shows is due to large values of dust attenuation (typically $A_V > 2$ mag even at large radii). Morphologically, the majority of our UFO sample tends to have low S\'ersic indices ($n \sim 1$) suggesting these large, massive, optically faint galaxies have little contribution from a bulge in F444W.  Further, a majority have axis-ratios between $0.2 < q < 0.4$, which Bayesian modeling suggests that their intrinsic shapes are consistent with being a mixture of inclined disks and prolate objects with little to no contribution from spheroids. While kinematic constraints will be needed to determine the true intrinsic shapes of UFOs, it is clear that an unexpected population of large, disky or prolate objects contributes significantly to the population of optically faint galaxies.

\end{abstract}

\keywords{galaxies: evolution -- galaxies: formation -- galaxies: high-redshift --galaxies: structure}

\section{Introduction} \label{sec:intro}

One of the more challenging populations of galaxies to characterize over the past few decades has been the optically dark/faint or ``HST-dark" galaxies, where strong dust attenuation causes a lack of UV-optical emission making them difficult or impossible to detect with facilities such as the Hubble Space Telescope (HST) \citep[e.g.,][]{barger1998,hughes1998,coppin2006,elbaz2011,williams2019,umehata2020,manning2022,barger2022,gomezguijarro2022a,xiao2023}. Surveys in the far-infrared (FIR) through millimeter wavelengths with instruments such as Spitzer/IRAC, Herschel, the James Clerk Maxwell Telescope (JCMT/SCUBA), and the Atacama Large Millimeter/Sub-mm Array (ALMA) have gradually revealed this population to be primarily concentrated around $z \gtrsim 2$  \citep[e.g.,][]{barger1998,hughes1998,coppin2006,elbaz2011,wang2012,wang2016,tadaki2017,franco2018,wang2019}. These and other sub-groups of optically faint and/or IR/sub-mm detected galaxies fall into the broad category of dusty star-forming galaxies (DSFGs), whose discovery, classification and properties are reviewed in \citet{casey2014}.  In addition to possessing large quantities of dust, measurements have revealed moderate to high stellar masses ($M_* \gtrsim 10^9 M_{\odot}$) and high star-formation rates \citep[SFRs, e.g.,][]{wang2019,elbaz2011} with a small subset undergoing intense bursts of star-formation with substantially higher SFRs of $\gtrsim 1000 / \mathrm{M_{\odot}/yr}$ \citep[e.g.,][]{walter2012,riechers2013,marrone2018,ma2019}. This obscured galaxy population appears to be an important, and possibly dominant contributor to the cosmic star-formation rate density out to $z \sim 4$ \citep[e.g.,][]{barger2012,madau_dickinson2014,zavala2021,barrufet2023}. 
	
Morphological studies of sources with detections at both optical/UV and sub-mm wavelengths have revealed that the DSFG population is often much more compact at longer wavelengths than at shorter wavelengths \citep[e.g.,][]{hodge2016,oteo2016,barro2016,rujopakarn2016,fujimoto2017,tadaki2017, nelson2019,gullberg2019}. However, this is not always the case, as \citet{sun2021} report two spatially extended DSFGs with sub-mm sizes exceeding the IR sizes from Spitzer/IRAC data. These two wavelength regimes, rest-frame optical and FIR, are commonly taken to represent the distribution of existing stellar mass and of star-formation, respectively, suggesting that star-formation is primarily centrally concentrated in these objects. This observation can be explained if these galaxies are in the process of building their central stellar bulges \citep[e.g.,][]{tadaki2017,nelson2019,tadaki2020}, a strong indicator of future cessation of star-formation \citep[e.g.,][]{franx2008,bell2012,lang2014,whitaker2017}. Lending further empirical evidence to support this physical interpretation, many DSFGs at higher redshifts ($z \sim 3$) have FIR sizes that are consistent with the optical sizes of quiescent galaxies at lower redshifts ($z \sim 2$). The leading hypothesis is thus that the DSFG population are the plausible progenitors of massive quenched systems at z $\sim$ 2 \citep[e.g.,][]{toft2014,fujimoto2017,tadaki2017,suess2021}, which in turn are likely the progenitors of the most massive and quenched galaxies observed in the local Universe. Given the relative importance of optically faint galaxies in the context of both the cosmic SFR budget and our understanding of massive galaxy evolution, it is imperative to have a more complete census of their morphologies and stellar populations \citep[e.g.,][]{gottumukkala2023,williams2024}.

Prior to the launch of the James Webb Space Telescope \citep[JWST,][]{gardner2023}, spatially-resolved studies of the optically faint population have primarily been facilitated by interferometers such as the Submillimeter Array (SMA) or ALMA \citep[e.g.,][]{walter2016,tadaki2017,cowie2018,franco2018,gomezguijarro2022a,chen2014,hsu2017}. Due to their limited field of view and lower sensitivity, studies have therefore been limited to bright, lensed, or previously known galaxies. Thus, while interferometers can achieve the necessary spatial resolution to characterize galaxies at higher redshifts, their small field-of-view makes it challenging to observe or discover large samples of galaxies. Conversely, infrared telescopes like Spitzer or Herschel can observe many galaxies, but their spatial resolution is too poor for morphological studies of galaxies. 

All of this has changed with the successful launch and commissioning of JWST, whose NIRCam instrument's large field of view ($\sim 10 \ \mathrm{arcmin^2}$) and wavelength coverage redder than HST have already enabled numerous studies of the optically faint population out to 4.4$\mu$m \citep{rieke2023}. 
Populations of these optically faint galaxies have been studied in the Cosmic Evolution Early Release Science (CEERS) survey \cite[e.g.,][]{barrufet2023,gomezguijarro2023} as well as in observations of the SMACS0723 cluster \citep{rodighiero2023} finding that these galaxies are generally at redshifts $2 < z \lesssim 7$ with high stellar masses ($M_* > 10^{10} M_{\odot}$) and high dust attenuations ($A_V > 2$). Breaking down the JWST-observed optically faint galaxies observed in CEERS, \citet{perezgonzalez2023} find that the majority of these sources (71\%) are DSFGs at $2 < z < 6$. However, to better understand how these galaxies fit into our overall picture of galaxy growth, we need to study their structures to better infer how these galaxies have assembled. \citet{gomezguijarro2023} study the stellar masses and morphologies of DSFGs at $3 < z < 7.5$ and find a subset of highly-attenuated (defined as $A_V > 1$) galaxies whose main difference from their total sample is their $\sim 30\%$ smaller effective size. 

Given the compact optically faint galaxies reported in \citet{gomezguijarro2023} and expectations that optically faint populations in general might be compact, the discovery of optically faint galaxies detected in CEERS with apparent sizes greater than 0.25” has come as a surprise \citep{nelson2023}. This visually striking galaxy population, dubbed Ultra-red Flattened Objects (UFOs), are noticeably more elongated and redder than the general population in both their inner and outer regions. Whereas the expectation may have been that these heavily obscured galaxies should be more compact than optically bright galaxies leading to a larger dust column density \citep[as in e.g.][]{gomezguijarro2023}, the UFOs are not more compact than other galaxies at the same mass. Given the low axis-ratios combined with inferred S\'ersic indices close to unity, \citet{nelson2023} suggest that it is most likely for the UFO population to be disk-galaxies. Interestingly, in a spatially resolved study of a single previously known and spectroscopically confirmed $z=2.38$ optically faint galaxy in the A2744 field, \citet{kokorev2023} are able to characterize the optically faint galaxy as being a massive ($\mathrm{log(M_*/M_{\odot}}) \approx 11.3$), highly star-forming ($200 \ \mathrm{M_{\odot}/yr}$), dusty, and edge-on spiral galaxy with a nearly uniform $A_V \sim 4$ across its disk. This galaxy has the same general characteristics of the UFO population described in \citet{nelson2023} and appears consistent with having a disk morphology based on the visual identification of spiral structure in F277W imaging. Although there is some evidence that UFOs are highly inclined disks, low S\'ersic indices and a projected axis ratio distribution skewed toward low values can also be indicative of an intrinsically prolate population. Existing samples so far have been too small to perform axis ratio modeling in order to statistically determine their intrinsic shapes, but here with deeper imaging over a wider area we can acquire large enough samples to perform statistically robust axis ratio modeling.

Looking at the observed color gradients (which can reveal spatial variations in stellar population properties, thus reflecting galaxy evolutionary and assembly histories) of the UFO population, \citet{nelson2023} find that they are consistent with previously observed negative color gradients in galaxies with colors becoming bluer as you move out from the center \citep[e.g.,][]{tortora2010,wuyts2010,guo2011,szomoru2013,chan2016,mosleh2017,suess2019a,suess2019b,miller2023}. However, the UFOs differ in that they are still red in their outskirts albeit not as red as in their inner regions. \citet{nelson2023} suggest that this could be driven by large quantities of dust in these objects. This dust would need to be distributed throughout most of a galaxy to account for the red colors throughout unless there were other spatial gradients in the UFO stellar populations (e.g., younger dust-obscured stars in the outskirts and older dust-obscured stars centrally). Resolved stellar population modelling is needed to determine why these objects are so red out to large radii.  

In order to better understand the intrinsic shapes of UFOs and to address what drives their observed colors, we present the identification and properties of 112 optically faint galaxies  found in the JWST Advanced Deep Extragalactic Survey (JADES) observations of both Great Observatories Origins Deep Survey (GOODS) fields. Of the 112 optically-faint galaxies, we identify 56 as being UFOs and, in this paper, we seek to understand the morphologies as well as the stellar population and dust content of these enigmatic objects leveraging the $\gtrsim 5$x larger sample along with deeper imaging. We describe our data and analysis methods in Section~\ref{sec:data} and present the results of the integrated stellar populations in Section~\ref{sec:stellar_pops}. In Section~\ref{sec:gradients}, we discuss the observed color gradients in the context of the stellar population and dust content of their inner and outer regions. We discuss the inferred projected shapes of the UFOs and show the constraints that can be made on their \textit{intrinsic} shapes in Section~\ref{sec:structure}. We discuss the implications of our results in Section~\ref{sec:disc} and we summarize our conclusions in Section~\ref{sec:concl}.

Throughout the paper, we assume the WMAP9 $\Lambda$CDM cosmology with $\Omega_m$ = 0.2865, $\Omega_{\Lambda}$ = 0.7135 and $\mathrm{H0 = 69.32 \ km s^{-1} Mpc^{-1}}$ \citep{hinshaw2013}. All magnitudes in this paper are expressed in the AB system \citep{oke1974}.

\bigspace

\section{Data Analysis} \label{sec:data}

\subsection{Observations}

JADES\footnote{Some of the data presented in this paper can be found in the Mikulski Archive for Space Telescopes (MAST),\dataset[10.17909/8tdj-8n28]{https://doi.org/10.17909/8tdj-8n28}} is a joint collaboration between the NIRCam and NIRSpec science teams carrying out observations of both GOODS fields with NIRCam and MIRI imaging as well as NIRSpec spectroscopy \citep{eisenstein2023}. Observations were conducted in 9 different filters (F090W, F115W, F150W, F200W, F277W, F335M, F356W, F410M, and F444W) spanning 0.90$\mu$m to 4.4$\mu$m and reaching flux limits of $~\sim$ 30 AB magnitudes in F090W, F115W, F150W, and F200W in the GOODS-S field whose data is mildly deeper than in GOODS-N ($\sim 1-1.5$ mags deeper than CEERS). Additional NIRCam medium-band data from the JWST Extragalactic Medium-band Survey \citep[JEMS; program 1963,][]{williams2023} is included. The areas covered by the GOODS-S and GOODS-N mosaics are $\sim$ 67 and $\sim$ 58 square arcminutes, respectively ($\sim 2$x wider than CEERS).

\subsection{Data Reduction and Photometry} \label{sec:phot}

We follow methods outlined in the first JADES data release \citep{rieke2023b} for the data reduction and photometry and briefly summarize the main steps here. The raw data was processed through the JWST Calibration Pipeline (v1.8.1, \citet{bushhouse2022}) using the CRDS pipeline mapping (pmap) context 1009 . The first two stages of the pipeline are followed using the default parameters to perform detector-level corrections as well as flat-fielding and the flux calibration. Before combining individual exposures, several custom corrections were performed to take care of features associated with the NIRCam images. Namely, these are a background subtraction using the \textsc{photutils} Background2D class, removal of 1/f noise associated with image readout, and subtracting the ``wisp" features using stacked wisp templates from JADES and other programs. Following astrometric alignment using a custom version of JWST TweakReg, the images for a given filter / visit are combined with Stage 3 of the JWST pipeline and these visit-level mosaics are then combined to produce the final mosaic (0.031"/pixel). 

Sources are detected following the methods in \citet{rieke2023b} using a signal-to-noise ratio (S/N) image generated from inverse-variance weighted stacks of the F277W, F335M, F356W, F410M, F444W science and error images as the signal and noise images, respectively. Photometry is performed for all of the sources in the catalog in circular apertures of diameters 0.2", 0.3", 0.5", 0.6", and an aperture enclosing 80$\%$ of the total energy. Errors are estimated by placing 100,000 random apertures in groups of 1000 across each band's mosaic and measuring the rms flux (in electrons) as a function of aperture size which is used to assess the contribution of the sky background to the total flux uncertainty. 

Our targets are selected using aperture-corrected, PSF-matched photometry with a circular aperture of diameter 0.3" and as such we infer the integrated stellar populations using the same photometry. For the purpose of determining the colors and the stellar populations of the inner and outer regions of each galaxy, we make use of the same inner aperture (without the aperture correction) to describe the inner region and use an outer annulus of width 0.15" (determined by subtracting the non-aperture corrected photometry within the inner 0.15" radius from the outer 0.3" radius) to describe the outer region. Additional Hubble Space Telescope (HST) photometry from the reductions of the Hubble Legacy Fields (HLF, \citealt{illingworth2016,whitaker2019}) is utilized in the SED fitting as well.

\subsection{Sample Selection}

\begin{figure*}
    \includegraphics[width=0.35\textwidth]{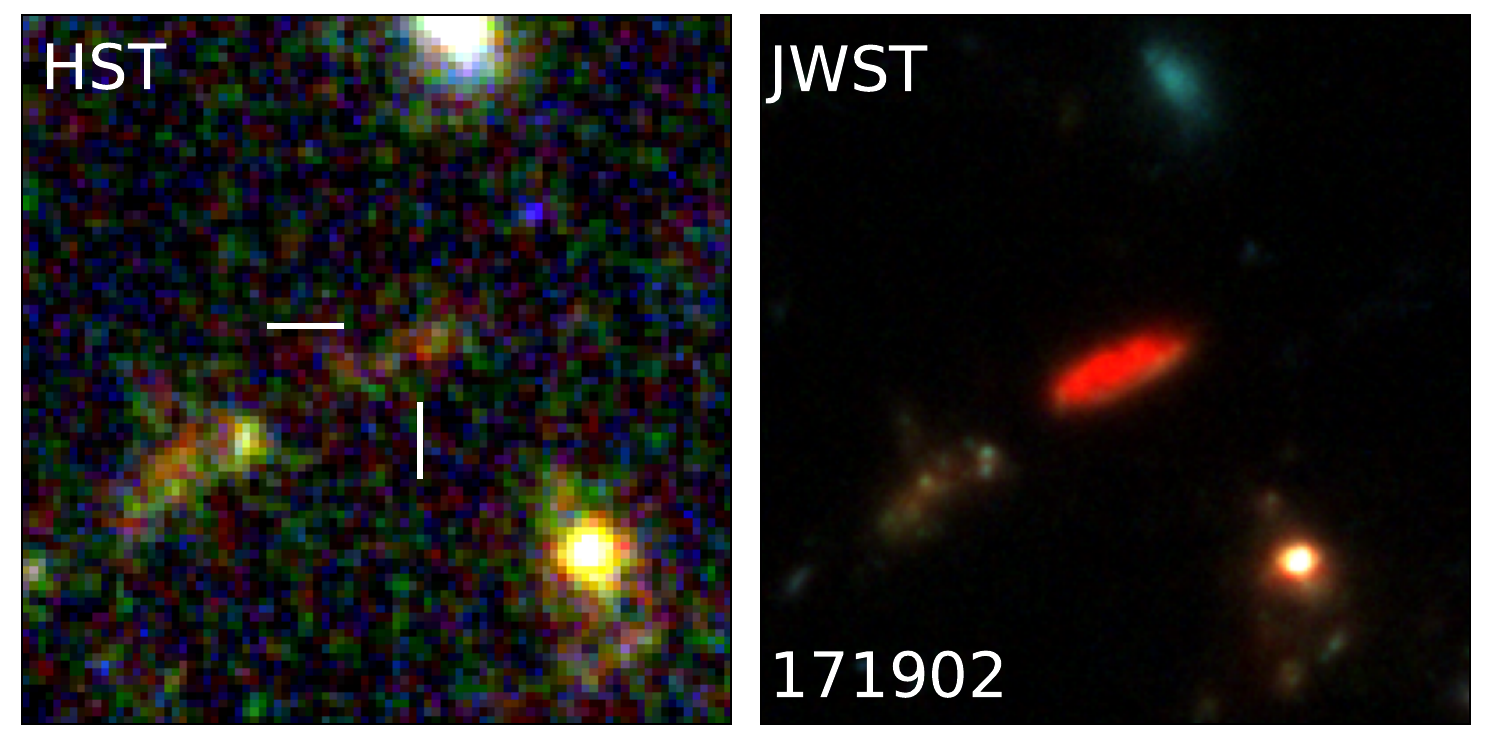}
    \includegraphics[width=0.35\textwidth]{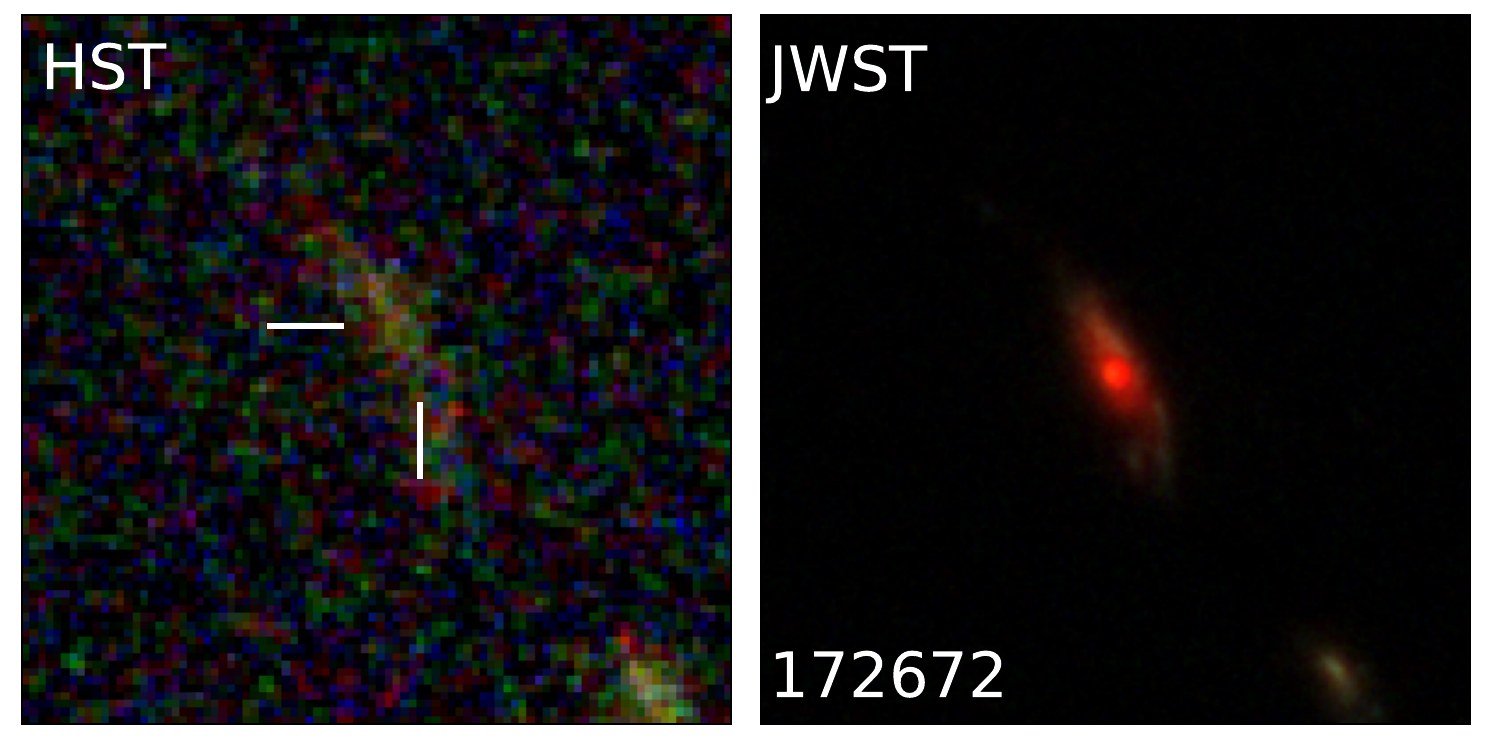}
    \includegraphics[width=0.35\textwidth]{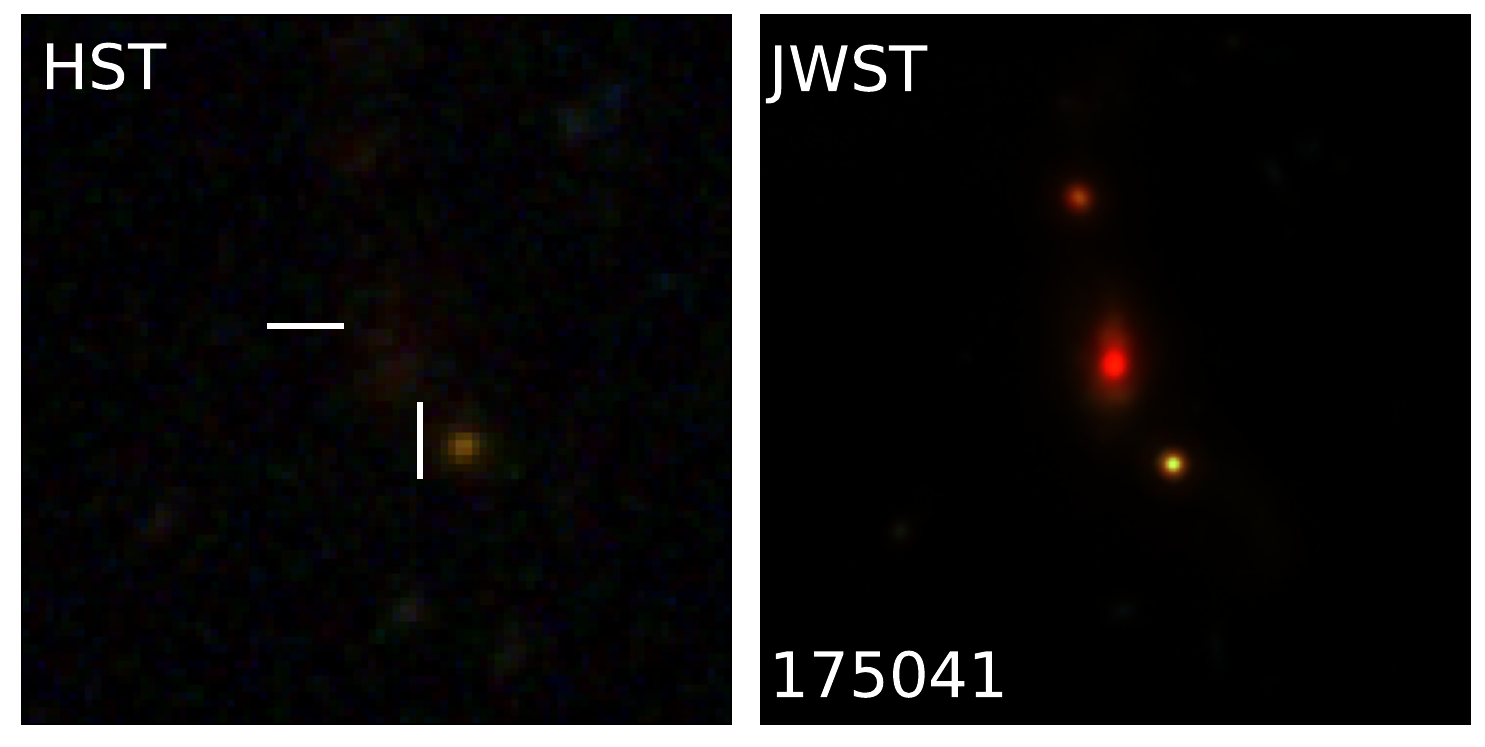}
    \includegraphics[width=0.35\textwidth]{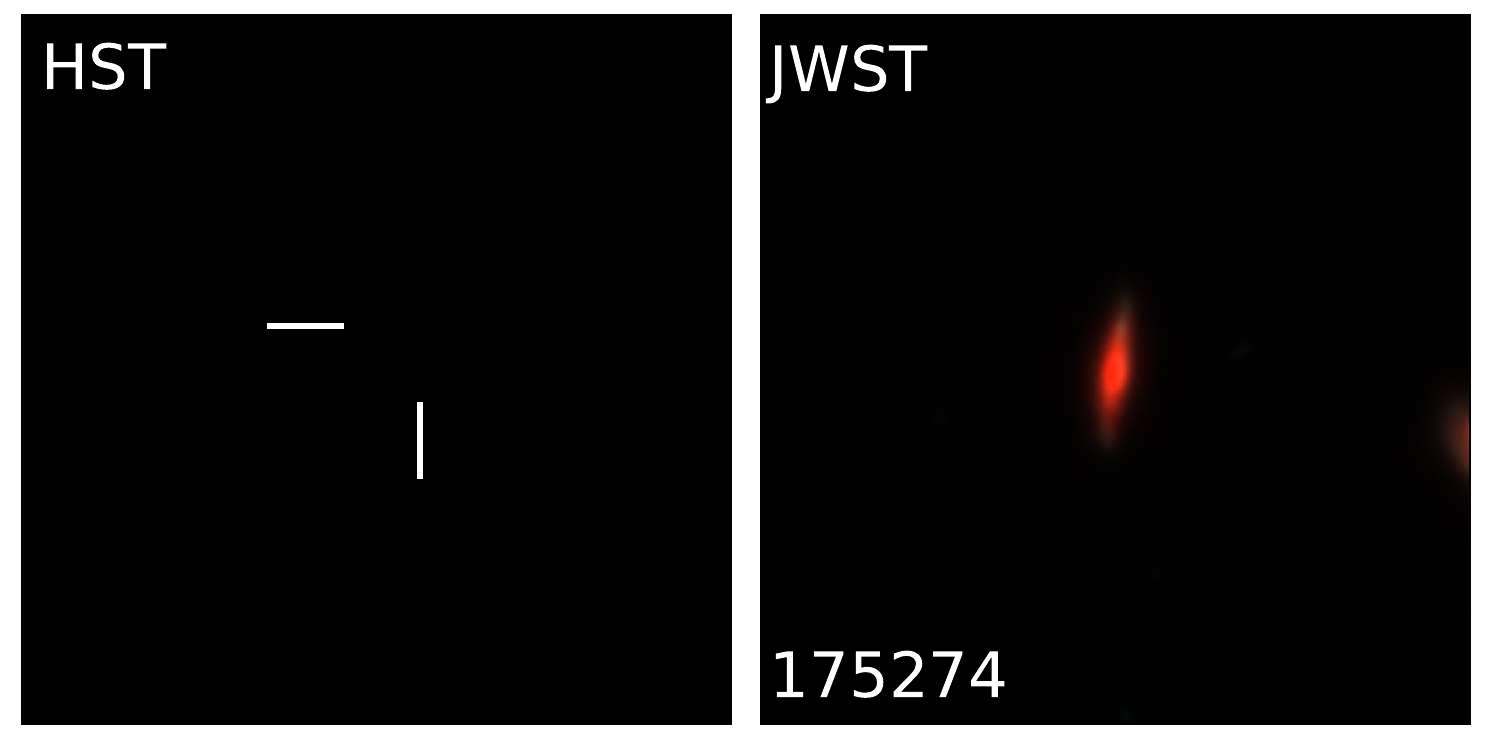}
    \includegraphics[width=0.35\textwidth]{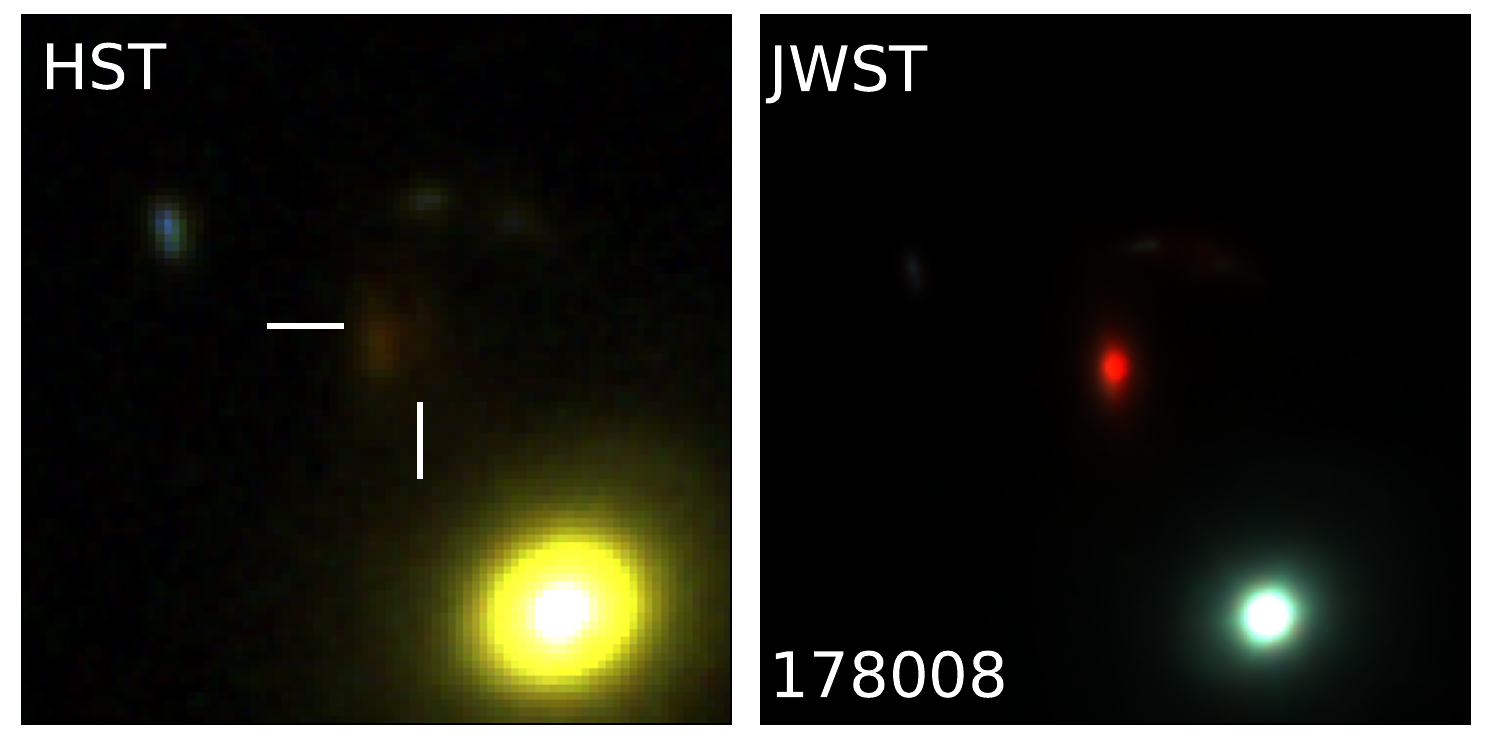}
    \includegraphics[width=0.35\textwidth]{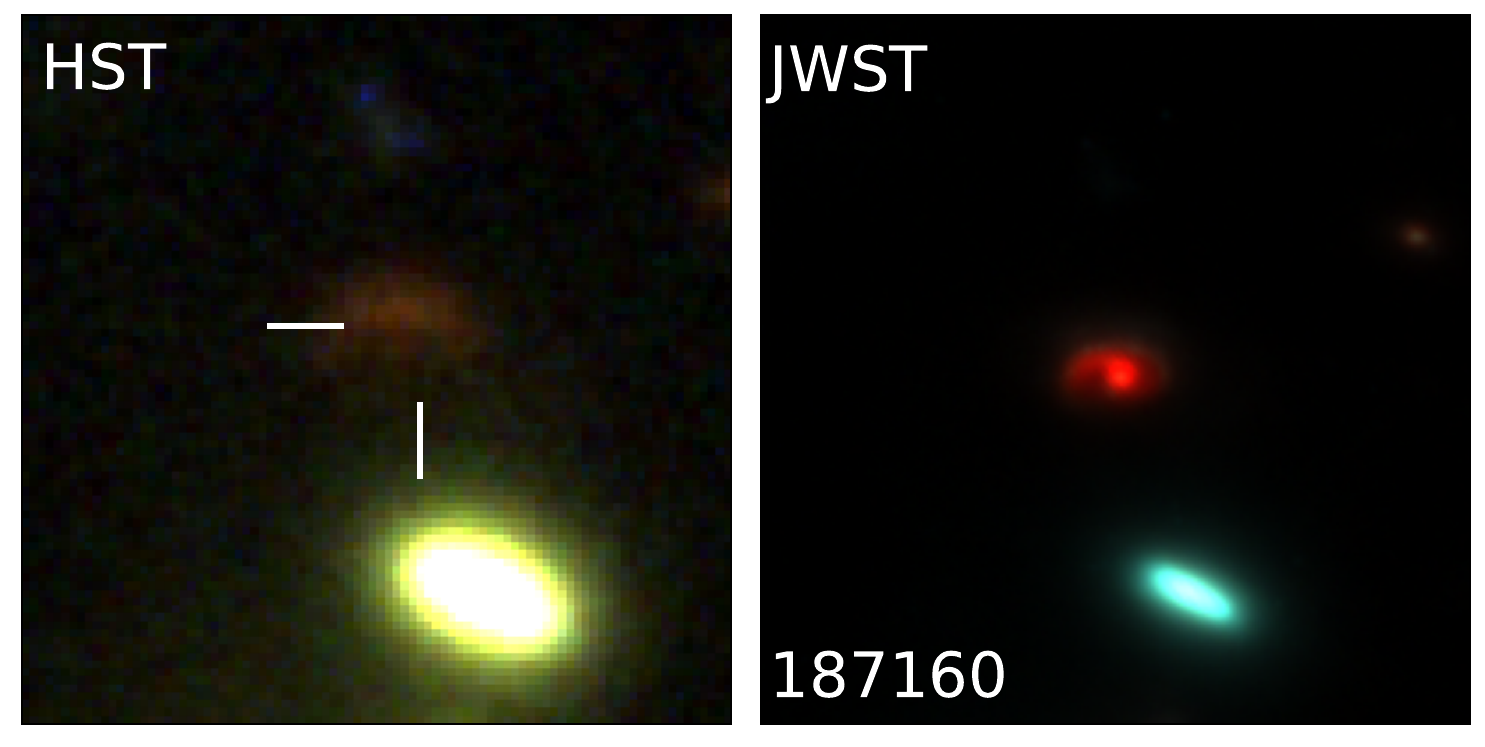}
    \includegraphics[width=0.35\textwidth]{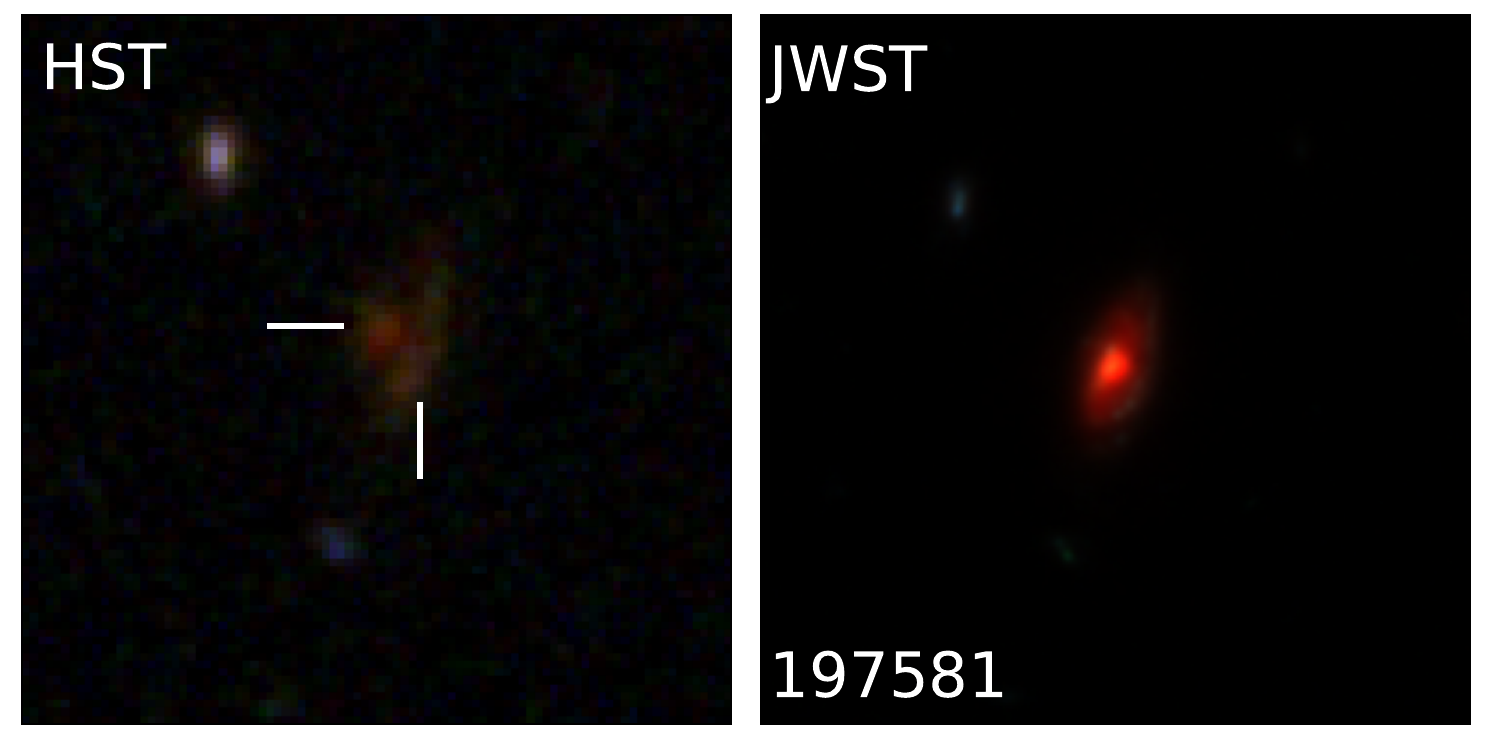}
    \includegraphics[width=0.35\textwidth]{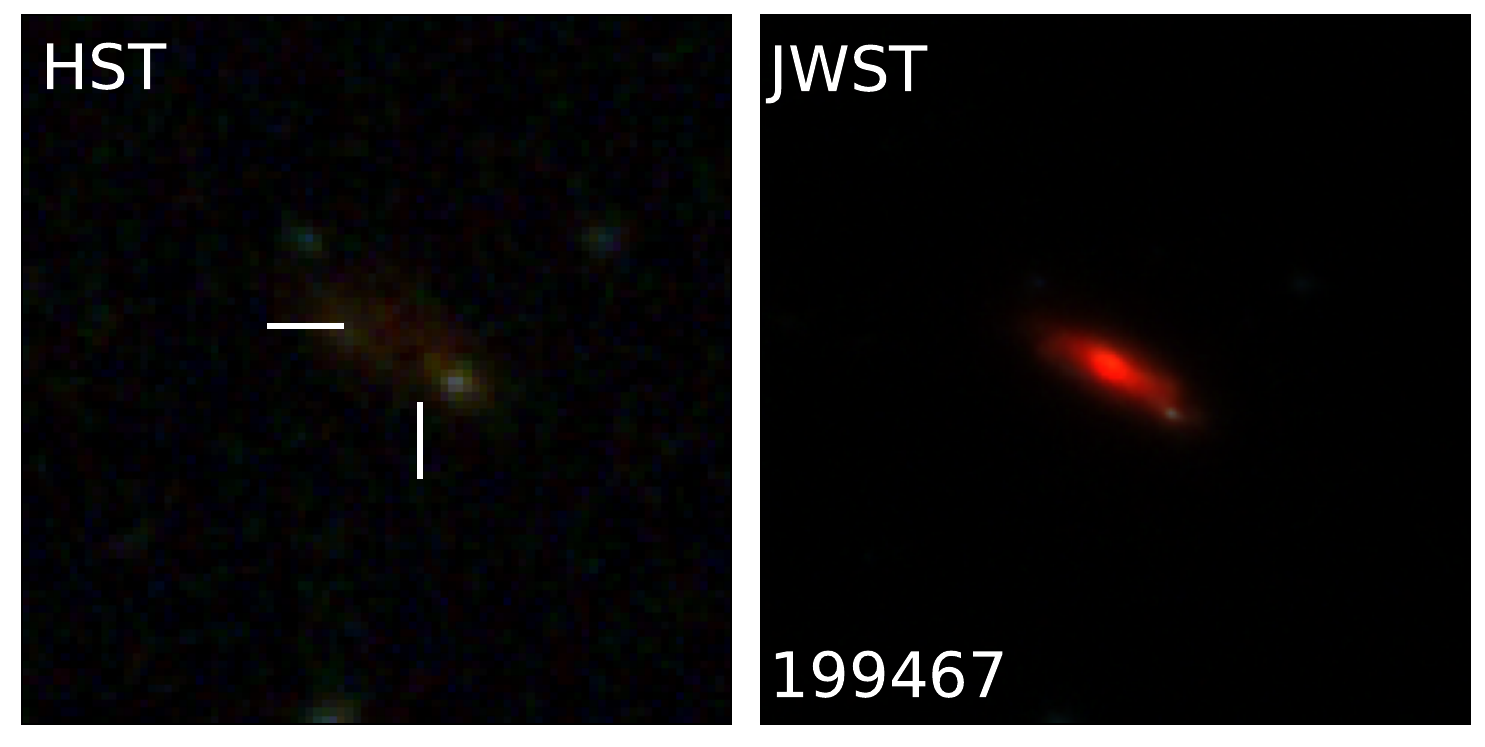}
    \includegraphics[width=0.35\textwidth]{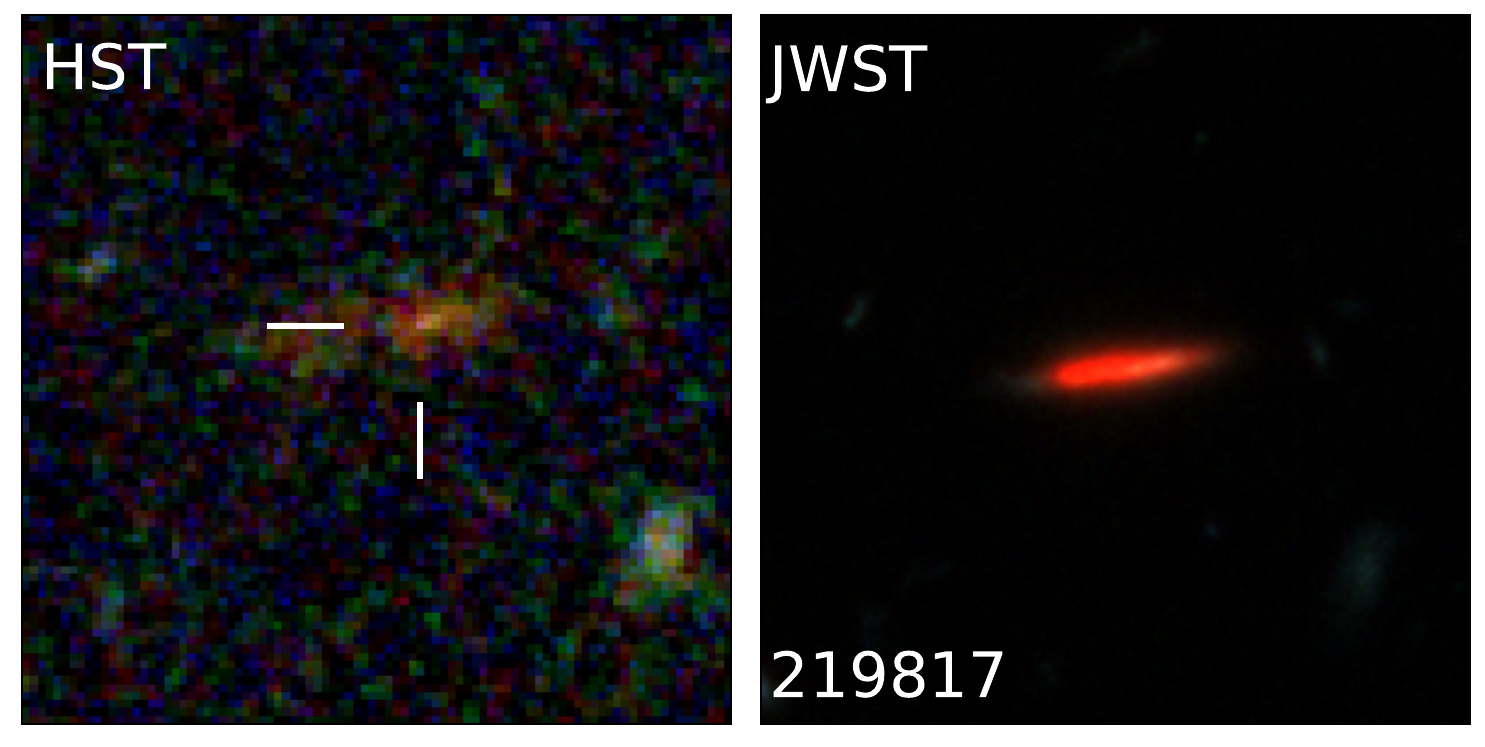}
    \caption{Comparison between the HST color image (F606W, F125W, F160W) and the JWST color image (F200W, F277W, F444W) for a representative sample of our UFOs. As can be seen, these galaxies are very bright and prominent with JWST imaging, but almost or completely invisible at HST wavelengths.}
    \label{fig:hst_v_jwst}
\end{figure*}

For this work, our goal is to leverage the larger area of the JADES survey to generate a larger sample of UFOs in an attempt to address their nature (e.g., are they disks?). To this end, we select the galaxies in JADES that are brightest at the reddest wavelengths of our coverage (4.4$\mu$m) and faint at HST wavelengths ($<$ 1.6$\mu$m). Examples of HST images for a selection of our targets can be seen in Figure~\ref{fig:hst_v_jwst}. From this sample of optically-faint galaxies, we take those with semi-major axis effective (half-light) sizes greater than 0.25" as our sample of UFOs.

Originally, we adopted the optically-faint galaxy (OFG) selection criteria in \citet{nelson2023} requiring AB magnitudes of F444W $<$ 24.5, F115W $>$ 27, and F150W $>$ 25.5. However, during our analysis we noted that some of the selected UFOs showed blue excesses either in the form of bluer clumps, excess blue emission on the outskirts, or SEDs more indicative of bluer stellar populations. These bluer outliers had inferred $A_V$ values much lower ($\lesssim 1$) than their redder companions, and thus to select a more intrinsically red parent sample from which to find UFOs, we revised the OFG selection criteria of \citet{nelson2023} to include the same F444W brightness cut, but with the addition of the following color cuts: F150W - F200W $>$ 0.75 and F200W - F444W $>$ 2.0. The selection change results in a much more robust sample of thoroughly red objects whose optical faintness is being largely driven by dust attenuation (see Section~\ref{sec:stellar_pops}). 

\begin{figure*}
    \centering
    \includegraphics[width=\linewidth]{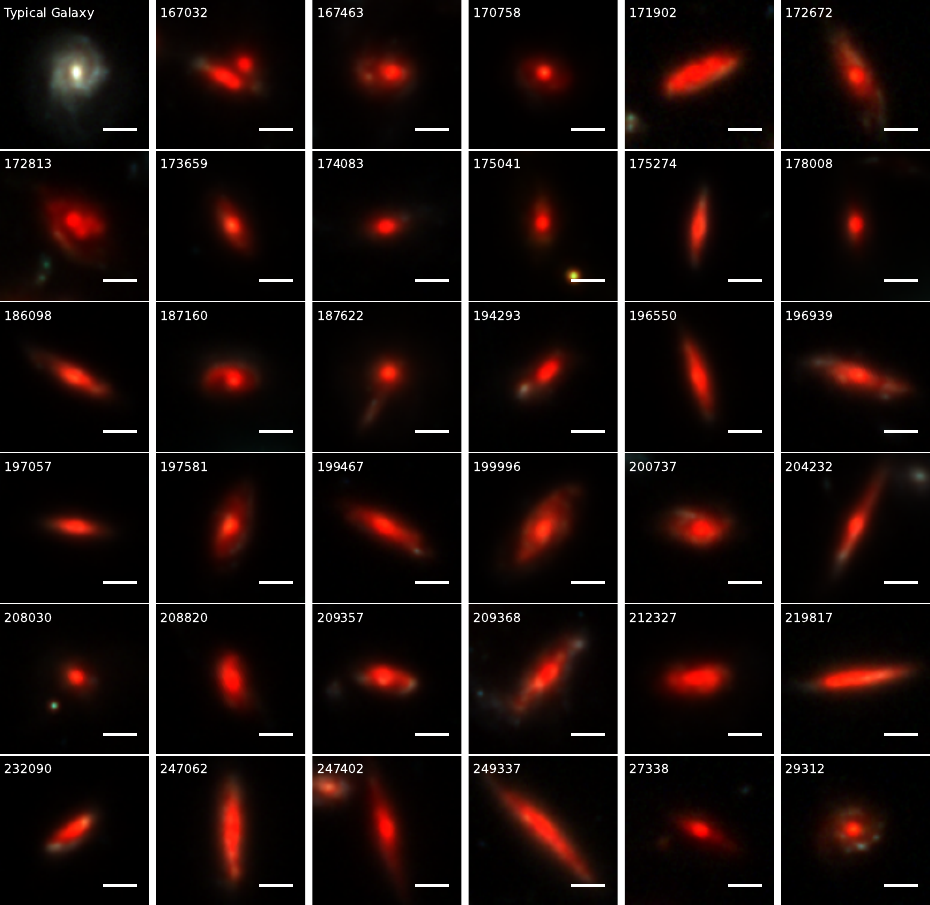}
    \caption{Color images (F200W, F277W, F444W) of 35 of our UFOs illustrating the extended red nature of these objects. Identifiers in the upper left are the IDs in the JADES photometric catalogues and a 1" bar is shown for reference in the lower right. These 1" scale-bars translate to physical sizes of 6.7 kpc to 8.5 kpc depending on the redshift of the object. For comparison, we show a ``typical" galaxy in the upper left panel with bluer colors and more pronounced color variations than the UFOs. }
    \label{fig:gallery}
\end{figure*}

With this criteria, we identify 112 total galaxies with a range of apparent sizes. We fit for their morphological parameters (described in Section~\ref{sec:morph_methods}) allowing us to identify UFOs amongst our sample of optically-faint galaxies. This selection identifies 56 UFOs whose stellar populations, morphologies, and comparisons with our broader optically-faint sample we discuss below. In Figure~\ref{fig:gallery} we show F200W, F277W, and F444W color images of the UFOs and in Figure~\ref{fig:hst_v_jwst} we show, for a sub-sample of UFOs, how HST color images (F606W, F125W, F160W) compares with the corresponding JWST image clearly illustrating the ability of JWST to reveal galaxies that were invisible or faint at HST wavelengths. 

\begin{figure}
    \centering
    \includegraphics[width=\linewidth]{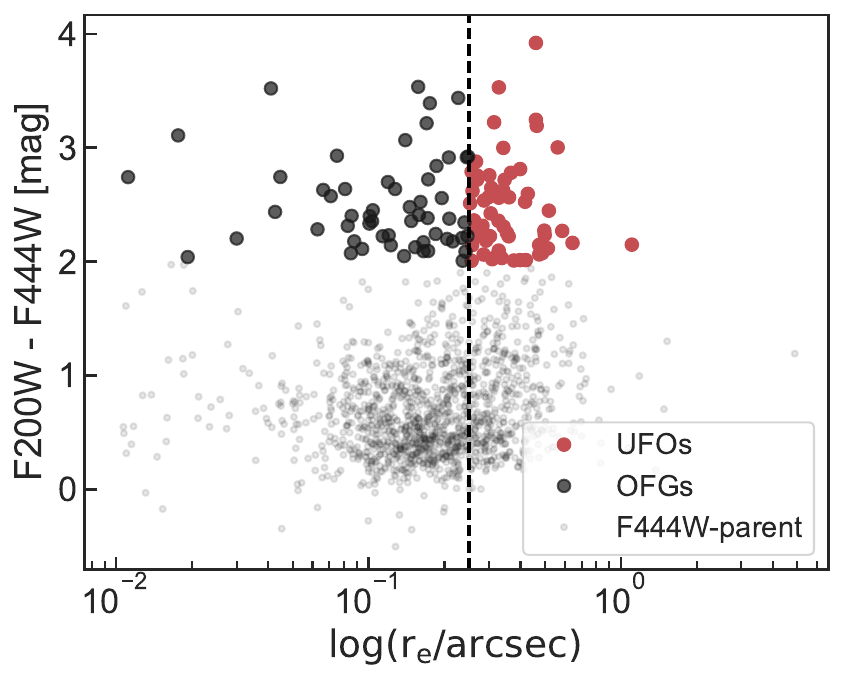}
    \caption{F200W-F444W color versus the semi-major axis effective size illustrating our optically-faint galaxy color selection and our UFO size cut (dashed black line at 0.25").  }
    \label{fig:sample_sel}
\end{figure}

Additionally, from the F444W $< 24.5$ mag parent sample, we identify those that are bluer than our optically faint selected sample in both colors and analyze them as well. This sample is further limited to the same redshift and stellar mass range as the whole optically-faint sample (described in Section~\ref{sec:stellar_pops}). We fit for the stellar populations and morphologies of this sample in the same way as for the optically-faint galaxies in order to more broadly compare the two populations. Unless otherwise stated, figures will refer to the blue sample of F444W-selected galaxies as the F444W-parent sample and the smaller (non-UFO) optically-faint galaxies as OFGs. Our three samples are shown in Figure~\ref{fig:sample_sel} where we plot their F200W - F444W colors versus their semi-major axis-effective size illustrating both our size cuts and one of our color cuts. 

\subsection{Stellar Populations} \label{sec:sp_methods}

We utilize the photometric redshift code, EAzY \citep{brammer2008}, to infer the redshifts of our sample of galaxies. The templates and photometry used to constrain the redshifts follow the methods outlined in \citet{hainline2023}. For the rest of the stellar population parameters, we use the \textsc{Prospector} SED fitting code \citep{johnson2021} fixing the redshift to the value inferred from EAzY. We verified that allowing the redshift to remain free in our \textsc{Prospector} fits yields redshifts consistent with EAzY as well as yielding consistent stellar population parameter inferences as the fixed-redshift results. To vastly decrease the time needed to perform each SED fit, we make use of a neural net emulator, \textsc{parrot}, that has been trained to infer the photometry for some underlying stellar population model (see \citet{mathews2023}). This emulator gives consistent results with standard \textsc{Prospector}, but with a factor of $\approx 10^3$  increase in speed.

We adopt the \textsc{Prospector}-$\beta$ model described in detail in \citet{wang2023} adjusted to keep the redshift fixed. In brief, this is the same model as the \textsc{Prospector}-${\alpha}$ model (\citealt{leja2017}), but makes use of a joint prior on redshift, stellar mass, and stellar metallicity disfavoring high mass, high-z solutions while still allowing a non-negligible probability of obtaining such solutions (thus making it possible to discover massive galaxies at high redshifts). The stellar populations are modeled using the MILES spectral templates \citep{sanchez-blazquez2006,falcon-barroso2011} with the MIST stellar isochrones \citep{choi2016} as implemented in FSPS \citep{conroy2009,conroy2010}. We adopt the Chabrier initial-mass function (IMF, \citet{chabrier2003}) and model the star-formation history (SFH) by fitting for the mass formed in seven logarithmically-spaced time bins using a continuity prior \citep{leja2019a} that weights against sharp changes in the SFR between adjacent time bins. In practice, this prior is a Student's-t distribution for the log ratio of the SFR in adjacent time bins, $\mathrm{log (SFR_n / SFR_{n-1})}$, with width of $\sigma = 0.3$, and degrees of freedom, $\nu = 2$. 

To accurately characterize the red and likely dusty nature of these galaxies, we adopt the two-component dust model of \citet{charlot2000} which characterizes the dust in galaxies with a diffuse dust component attenuating all stars equally and a separate birth-cloud component providing additional dust attenuation for stars younger than 10 Myr. Each component can have extinctions up to $A_V \sim 4.3$ and we also allow for a varying dust attenuation curve following the prescription in \citet{noll2009}. We additionally test two alternative dust models in which we: 1) remove the young star component, and 2) fix the slope of the attenuation curve to the \citet{calzetti2000} value. These two variations give consistent results with our fiducial dust model. The effects of dust emission are included using the dust emission templates from \citet{draine2007} with three free parameters that control the shape of the infrared (IR) SED (see \citet{leja2017} for details). Finally, the possibility of an AGN powering hot dust emission is included \citep{leja2018} utilizing the AGN templates of \citet{nenkova2008a,nenkova2008b} and  nebular line emission is included following the \textsc{Cloudy} \citep{ferland1998,ferland2017} FSPS implementation described in \citet{byler2017}. We note that certain aspects of this model (i.e., far-infrared dust and AGN emission) are not strictly necessary, but are included as default parameters in the emulator we use. It is possible to fix certain parameters to narrow the explored parameter space, but doing so has negligible impact on our results. 

\begin{figure*}
    \includegraphics[width=0.33\textwidth]{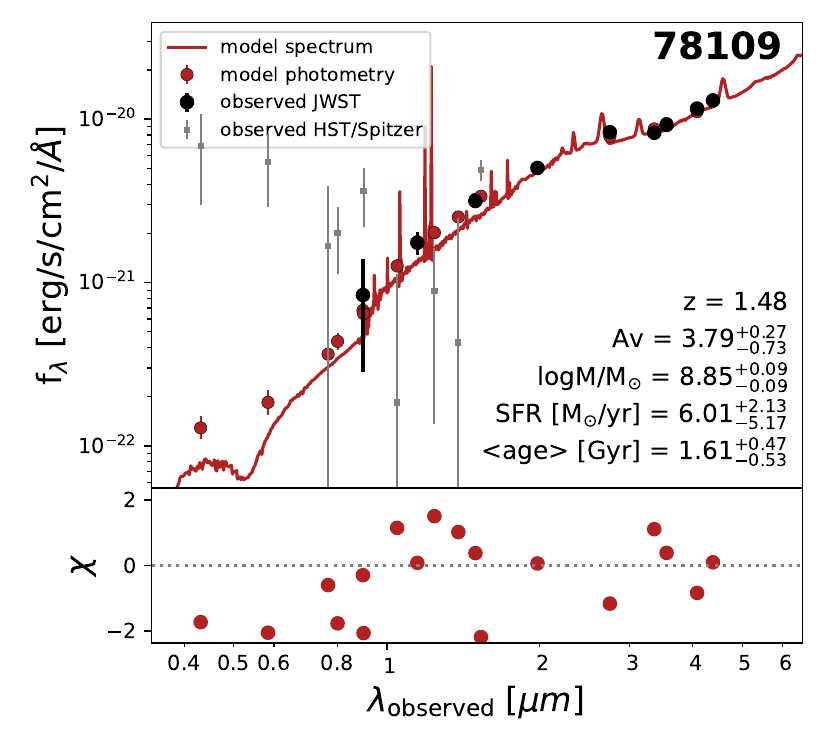}
    \includegraphics[width=0.33\textwidth]{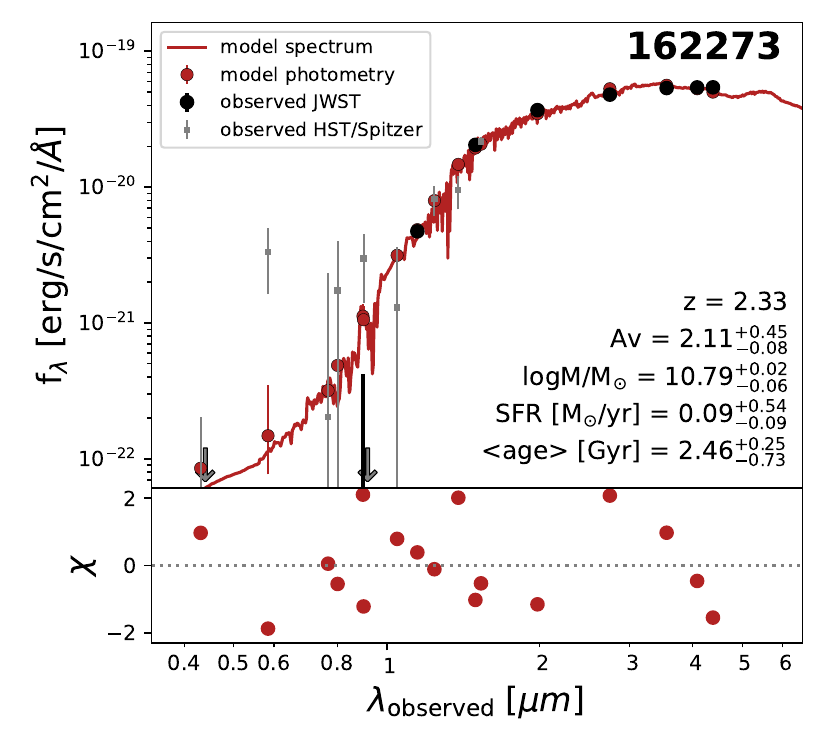}
    \includegraphics[width=0.33\textwidth]{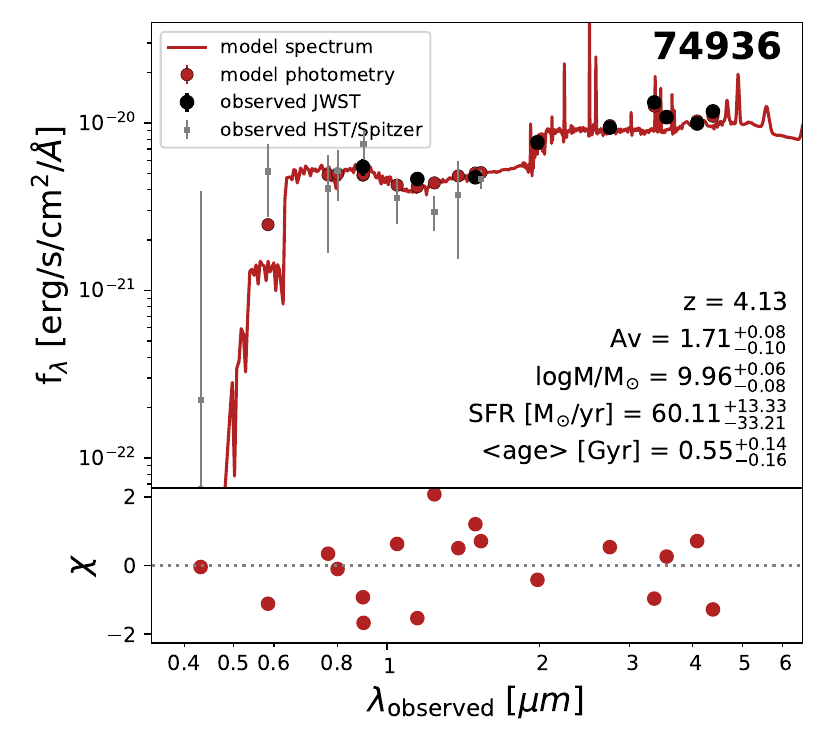}
    \includegraphics[width=0.33\textwidth]{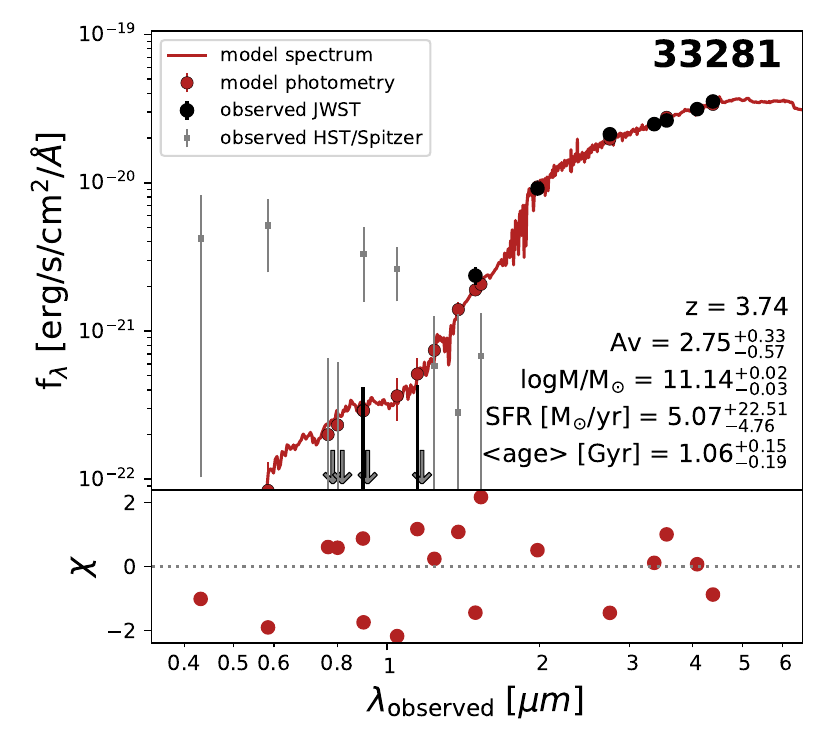}
    \includegraphics[width=0.33\textwidth]{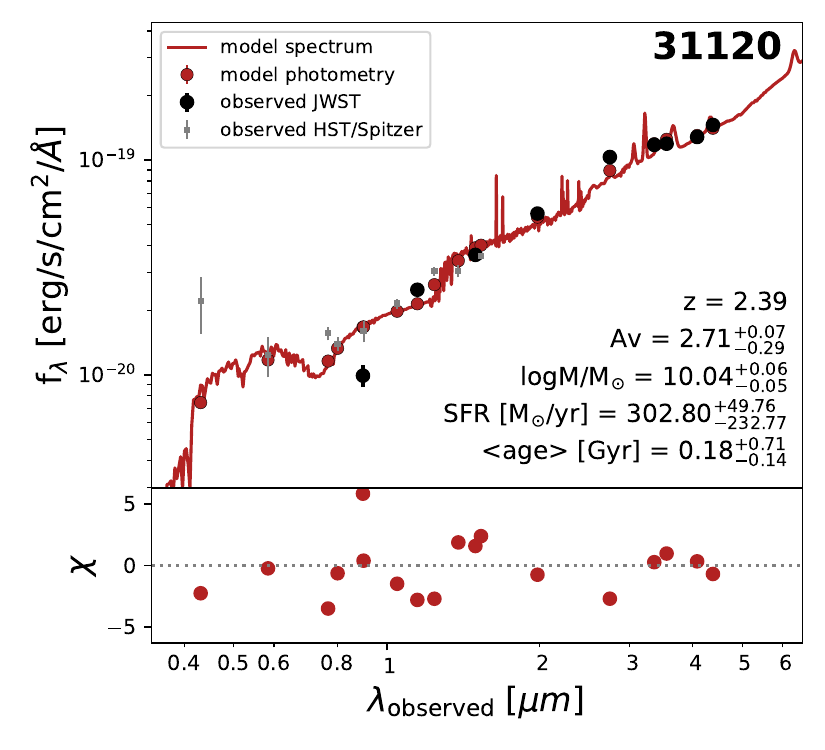}
    \includegraphics[width=0.33\textwidth]{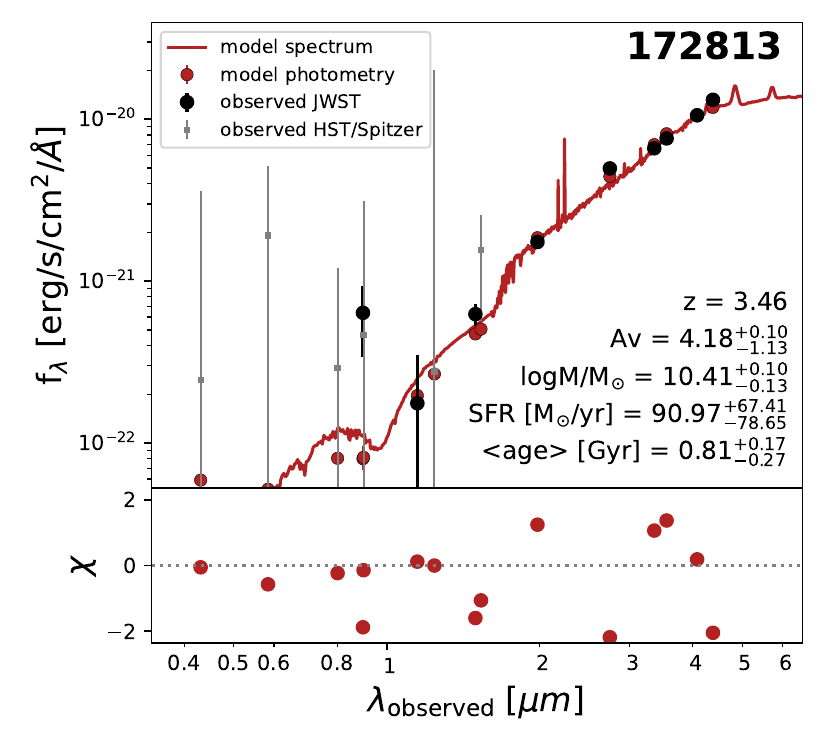}
    \caption{Observed and model SEDs for a representative sample of our UFOs with the observed JWST photometry as black points, observed HST/Spitzer photometry as gray points, model photometry as red points, and the model spectrum in red. Photometric points with negative fluxes are indicated with the gray arrows. The SEDs in each column show the lowest (top SED) and highest (bottom SED) value galaxy in terms of stellar mass (left), sSFR (middle), and diffuse dust $A_V$ (right). We note that some of the SED fits are not ideal, particularly for the HST/Spitzer photometry of the galaxies in the left column. However, fits like these are  not representative of the majority of our sample, and we also find that the stellar population parameter inferences are not strongly changed if only JWST photometry is used, so we keep these objects in our analysis}.
    \label{fig:seds}
\end{figure*}

We fit each object using the photometry as described in Section~\ref{sec:phot} and values are reported as the median of the posteriors and 1-$\sigma$ uncertainties as the the 84th–50th and 50th–16th interquartile ranges. To approximately account for systematic errors or issues with early JWST photometric calibrations, we enforce an error floor of 5\%. We show the results of our fits in Figure~\ref{fig:seds} for a representative group of galaxies in our sample. Each fit shows the observed SED, the model SED, and a spectrum generated in FSPS using the inferred model parameters. 

\subsection{Morphologies} \label{sec:morph_methods}

We fit for the semi-major axis effective (half-light) sizes and other morphological parameters (S\'ersic indices, $n$, and axis-ratios, $q=b/a$) for the entire sample of optically-faint galaxies with \textsc{galfit} \citep{peng2002,peng2010} and with \textsc{Lenstronomy}\citep{birrer2018,birrer2021}. The setup we use for \textsc{galfit} follows that of \citet{suess2022b} using empirical point-spread functions (ePSFs) generated with the \textsc{EPSFbuilder} class in \textsc{photutils} (\citealt{ji2023,anderson2000,anderson2016}). From the mosaics, we create 80x80 pixel cutouts centered on each object which sufficiently captures all of the light from each target. Sources and masks in each cutout are defined using \textsc{photutils} with any object within 3" of the target center and no more than 2.5 mag fainter than the target being fitted.  Finally, we subtract the background from each image using the SExtractor algorithm in \textsc{photutils} and fit for the morphological parameters in both the F200W and F444W filters. 

\textsc{Lenstronomy} is designed for the modeling of  gravitational lenses, but is also equipped to perform image modeling in a similar way as \textsc{galfit} (i.e., via S\'ersic profile fitting taking into account the PSF). By fitting with both \textsc{Lenstronomy} and \textsc{galfit}, we obtain a rough way to validate the morphological fits with the two fits tending to give consistent results with few exceptions. The two codes return similar results for the three morphological parameters ($n$, $q$, $r_e$), with median differences between the two fits of less than 0.1 for each parameter. Additionally, as discussed in Section~\ref{sec:disc}, we perform modeling of the observed axis-ratio distribution to constrain the UFOs intrinsic shapes and by using two sets of results we can obtain a rough approximation of the median uncertainty on the axis-ratio measurements. Given that \textsc{Lenstronomy} is able to fit the entire sample of UFOs and OFGs whereas \textsc{galfit} fails on multiple objects, we adopt the \textsc{Lenstronomy} results as our fiducial morphological parameter values. 

\section{Integrated Stellar Populations} \label{sec:stellar_pops}

\begin{figure*}
    \includegraphics[width=0.5\textwidth]{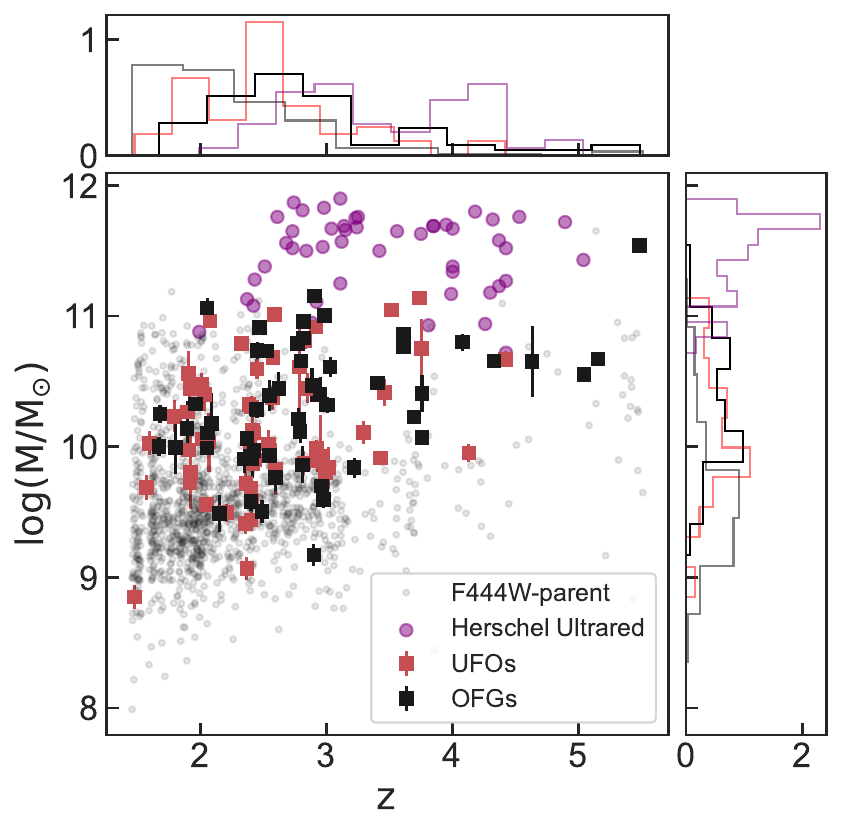}
    \includegraphics[width=0.478\textwidth]{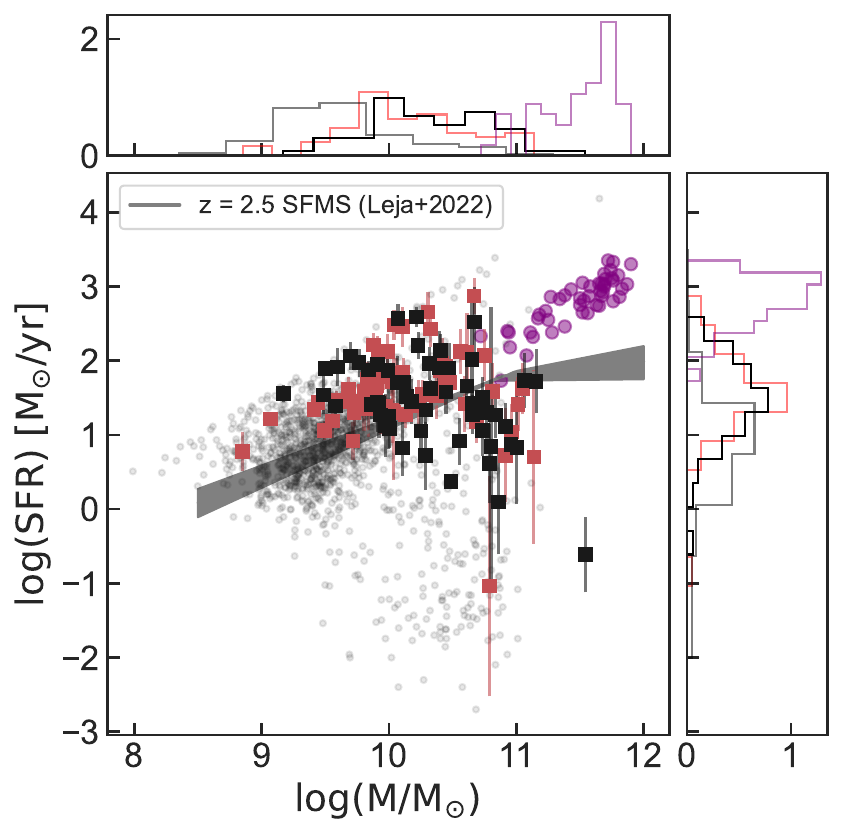}
    \caption{Left panel: Stellar mass inferred from SED fitting versus the redshift inferred from EaZY with UFOs in red, smaller optically-faint galaxies in black, the bluer F444W-selected sample in gray, and ultrared SMGs from \citet{ma2019} in purple. Right panel: how our objects are distributed in SFR and $\mathrm{log(M/M_{\odot})}$ with the shaded gray region showing the SFMS at $z = 2.5$ from \citet{leja2022}. The UFOs are a much less extreme population than the optically-faint FIR-selected galaxies, with lower stellar masses and SFRs. They occupy the high mass end of all F444W-selected galaxies showing that it is not only extreme star-bursts that were missed in previous censuses, but also more moderate galaxies. 
    }
    \label{fig:stellar_pop_comp}
\end{figure*}

In Figure~\ref{fig:stellar_pop_comp}, we show how our galaxies are distributed in stellar mass and redshift as well as how they are situated within the “star-forming main sequence” (SFMS) of \citet{leja2022} for $z=2.5$ galaxies. The UFOs tend to occupy the same region of both the $z-M_*$ and $SFR-M_*$ plane as their optically faint parent sample suggesting that the primary distinguishing factor is their structure. These objects tend to be clustered between $2 < z < 4$, with moderate to high stellar masses ($\mathrm{log(M_*/M_{\odot})} > 10$), and SFRs that tend to lie on or above the SFMS of $z=2.5$ galaxies from \citet{leja2022} except at the massive end where there is a noticeable drop to below the SFMS at a mass of $\sim 10^{10} \mathrm{M_{\odot}}$. We discuss this drop more in Section~\ref{sec:disc}. Median values and percentiles (16th and 84th) for these three parameters ($\mathrm{z, log(M_*/M_{\odot}), log(SFR/(M_{\odot}/yr)}$) are $2.42^{+0.58}_{-0.50}$, $10.04^{+0.63}_{-0.33}$, and $1.58^{+0.50}_{-0.37}$ for the UFOs and $2.78^{+0.85}_{-0.64}$, $10.29^{+0.50}_{-0.39}$, and $1.48^{+0.48}_{-0.57}$ for the smaller optically faint galaxies. We show these values along with the values for other populations in Table~\ref{tab:int_stellar_pops}.

Fig.~\ref{fig:stellar_pop_comp} also compares the optically faint populations in the context of the bluer F444W-selected parent sample and a FIR-selected sample from \citet{ma2019}. The FIR-selected galaxies are much more extreme than the UFOs with stellar masses predominately higher than $M_{*} \sim 10^{11} M_{\odot}$ and SFRs $> 100 \mathrm{M_{\odot}/year}$, and $z>3$, although we note that there is likely a systematic offset in their measured SFRs towards higher values compared to those we infer here owing to different methodologies used \citep[e.g.,][]{leja2019b}. These galaxies are examples of extremely red objects with large $A_V$ that were selected to have rising Herschel/SPIRE flux densities ($S_{500} > S_{350} > S_{250}$). Compared to the bluer F444W-selected galaxies, the optically-faint galaxies tend to have slightly higher redshifts and stellar masses than the optically-faint galaxies. The SFMS in \citet{leja2022} is fit to the 1.6\micron-selected 3D-HST galaxies, so the fact that the majority of our sample lies on and around this line means that these are not extreme objects but rather a normal subset of the massive galaxy population that was previously missed. Thus, the population of UFOs studied here are not extreme objects in regard to stellar mass, star formation rate, or redshift.  

\begin{figure*}
    \includegraphics[width=0.5\textwidth]{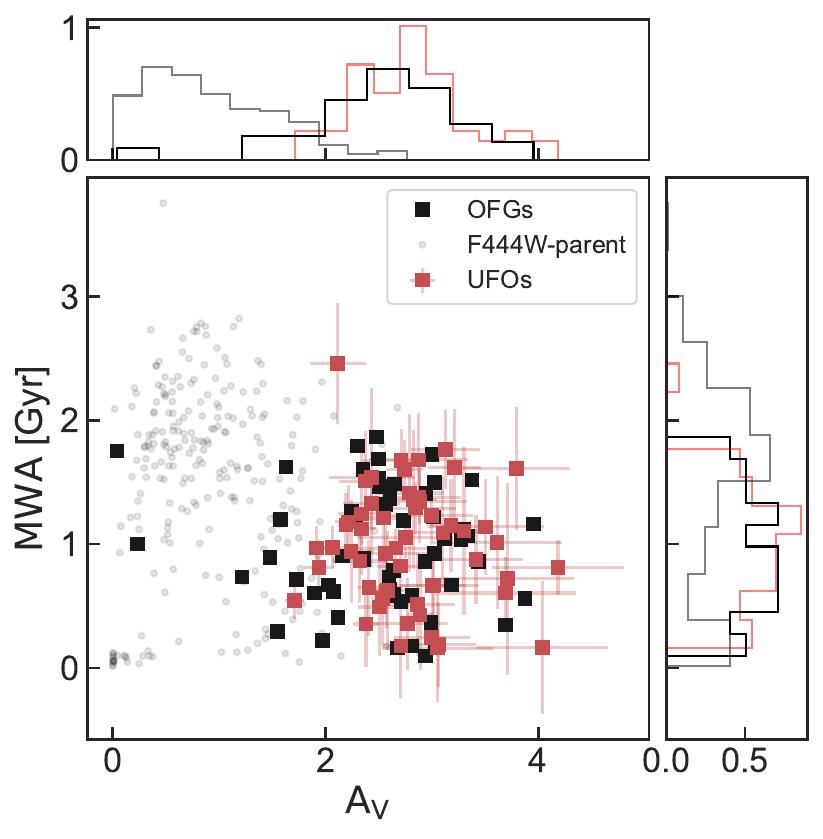}
    \includegraphics[width=0.5\textwidth]{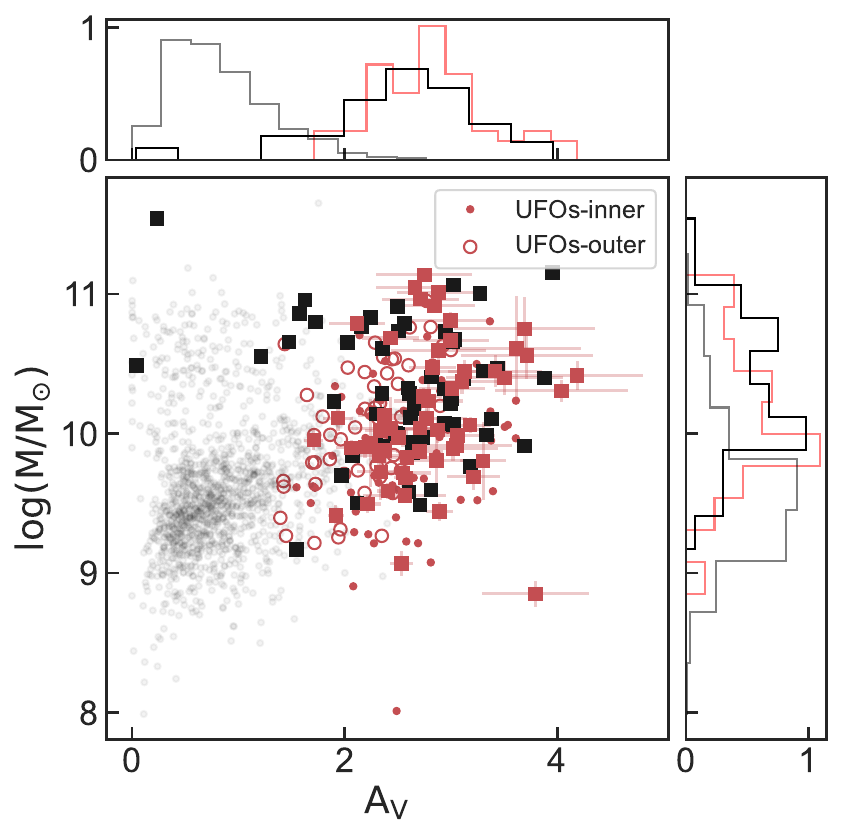}
    \caption{Left: mass-weighted age (MWA) versus dust attenuation for the UFOs (red), smaller optically faint galaxies (black), and the bluer F444W-selected galaxies in gray. The optically faint galaxies have larger dust attenuations than the bluer comparison sample, but similar MWAs suggesting that the redness of the optically faint galaxies is largely driven by elevated dust attenuation. Right: stellar mass versus dust attenuation for the same samples along with the values inferred for the inner and outer regions of the UFOs as described in Section~\ref{sec:gradients}.
    }
    \label{fig:mwa_av}
\end{figure*}

We also show the position of the three F444W-selected samples in the mass-weighted-age (MWA)$-A_V$ plane and the stellar-mass $A_V$ plane in Figure~\ref{fig:mwa_av} with the MWA and $A_V$ (for the diffuse dust component) representing two likely drivers of optical faintness. Most of the optically faint galaxies have large $A_V$ ($2.76^{+0.47}_{-0.43}$ for UFOs and $2.61^{+0.52}_{-0.65}$ for the smaller OFGs) and a young MWA ($0.97^{+1.43}_{-0.51}$ for UFOs and $0.90^{+1.50}_{-0.51}$ for OFGs). The right panel additionally shows the inferred $A_V$ in the inner and outer regions as discussed in Section~\ref{sec:gradients} demonstrating the large $A_V$ values in both the inner and outer regions of the UFOs. Compared to the bluer F444W-selected galaxies, the UFOs have much higher dust attenuation, but similar mass-weighted-ages suggesting that their red colors -- and the reason they were previously missed is owing to dust attenuation and redshift as opposed to old stellar ages. These results are discussed further in Section~\ref{sec:disc}.

To further quantify the differences between these populations (UFOs, smaller OFGs, and F444w-parent) in these 2D parameter spaces, we perform 2D Kolmogrov Smirnov (KS) tests on these distributions, using the public code \textsc{ndtest}\footnote{\url{https://github.com/syrte/ndtest}}, based on the algorithm described in \citet{Peacock1983} and \citet{Fasano1987}. First, in comparing the F444W-parent sample with either optically-faint population, we find very small p-values ($\mathrm{< 10^{-4}}$) showing that the optically-faint galaxies are a distinct population in terms of stellar mass, SFR, dust attenuation, and age. Comparing the UFOs and the smaller OFGs, we find a large p-value (0.53) for the MWA-$A_V$ distribution, a moderate p-value (0.14) for the log(SFR)-$\mathrm{log(M/M_{\odot})}$, and small p-values ($< 0.06$) for the z-$\mathrm{log(M/M_{\odot})}$ and $A_V$-$\mathrm{log(M/M_{\odot})}$ distributions. This suggests that UFOs are most similar to their smaller counterparts in their ages, dust attenuation, and SFRs, while stellar mass is where UFOs differ most from the smaller OFGs.

\begin{center}
\input{int_stellar_pop_props}

\end{center}

\section{Radial patterns in colors and stellar populations} \label{sec:gradients}

\begin{figure}
    \includegraphics[width=0.5\textwidth]{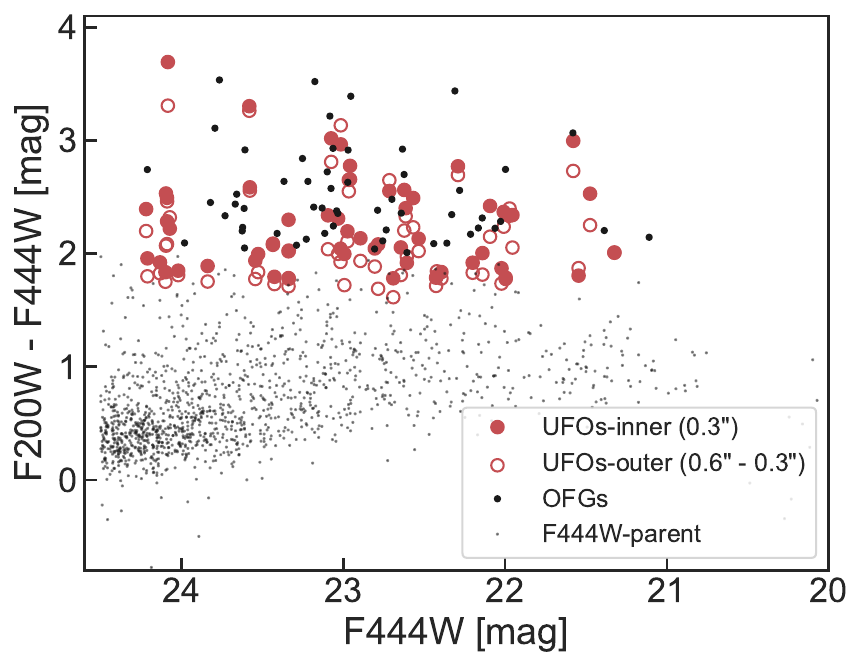}
    \caption{Observed colors versus observed AB magnitude in F444W for our optically faint galaxies and the bluer, mass and redshift matched, F444W-selected sample. For the UFOs, we show the color in the inner 0.3" circular aperture (filled red circles) as well as in an outer annulus of width 0.3" (open red circles). The optically-faint galaxies are clearly much redder than the parent sample with UFOs being consistently red throughout with red colors in their inner and outer regions. }
    \label{fig:color_grads}
\end{figure}

Looking at the images of our galaxies, it is apparent that these galaxies are consistently red throughout their extended light profiles. To make this visual assessment more quantitative, we calculate the F200W-F444W colors for the UFOs in the inner regions (0.3" diameter circular aperture) and outer regions (0.6" - 0.3" diameter apertures) of the PSF-matched images. We additionally calculate the integrated colors for the smaller optically faint and bluer F444W-selected galaxies. We show these colors in Figure~\ref{fig:color_grads} with the UFOs inner and outer colors shown with filled and unfilled red circles, respectively. As can be seen, the UFOs are strikingly red in both their inner and outer regions with both regions colors much redder than the bulk of F444W-selected galaxies. These are the same trends that were observed in the sample of UFOs first reported in \citet{nelson2023} where they saw that their 12 galaxies had red colors throughout the inner and outer regions that were generally much redder than the parent population. This is a unique population in that the inner and outer regions are both truly red as opposed to many galaxies that are observed to have red centers and blue outskirts. However, the color gradient is still negative in both cases with the UFOs being less red in the outskirts than they are in the centers. 

\begin{figure*}
    \includegraphics[width=1\textwidth]{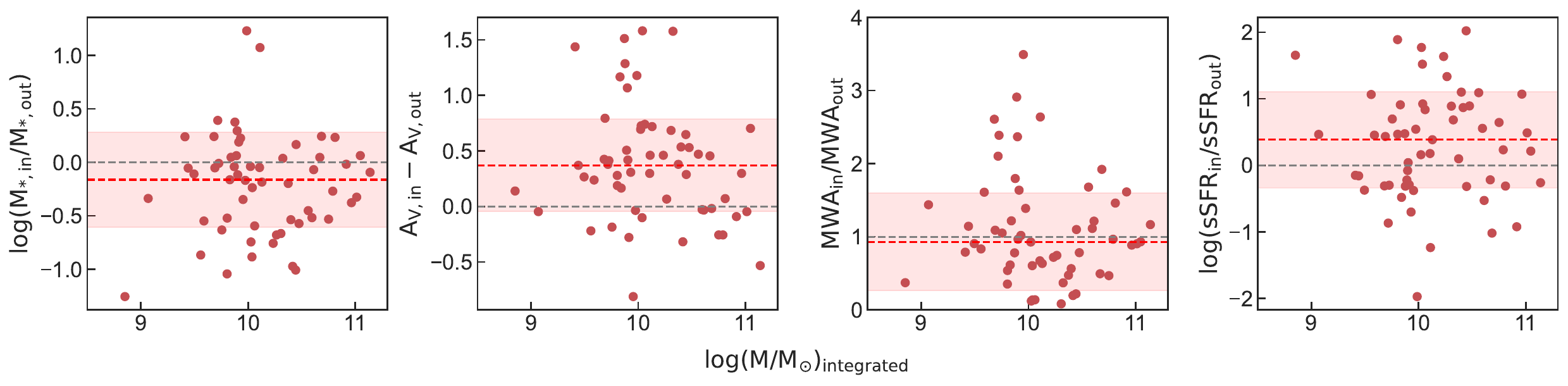}
    \includegraphics[width=0.9\textwidth]{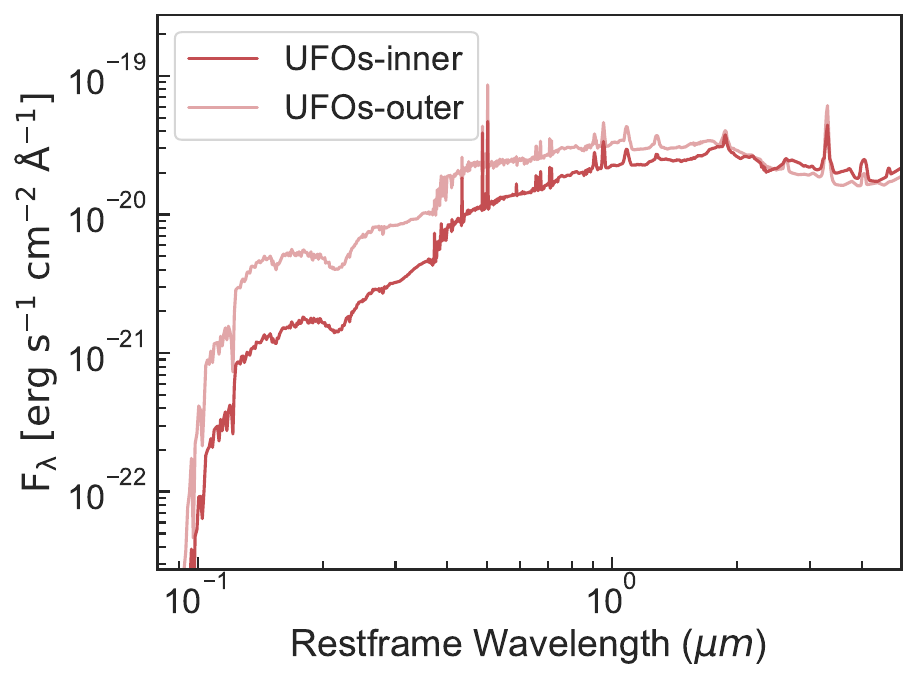}
    \caption{How the inferred stellar population parameters and predicted spectra of the UFOs compare between their inner regions (0.3" circular aperture) and outer regions (0.6" - 0.3" annulus). Top Panels: From left-to-right we show the ratio or difference between the inner and outer stellar mass, $A_V$, MWA, and sSFR as a function of the stellar mass inferred from the integrated fits discussed in Section~\ref{sec:stellar_pops}. The red and gray dashed lines show the median value of the ratio and where the values are equal. The red stripe shows the $\sim 1\sigma$ spread in the ratio or difference. Bottom Panel: median posterior spectra for the inner and outer regions generated from the maximum probability samples of the individual objects showing that the inner and outer median UFO spectra are most similar at redder wavelengths.}
    \label{fig:resolved_sps}
\end{figure*}

To further elucidate the persistent redness shown in Figure~\ref{fig:color_grads}, we also infer the UFO stellar populations within the inner 0.3" diameter (2.42 kpc at $z=2.5$) and the outer 0.6" - 0.3" annulus. This approximate spatially resolved analysis serves to place constraints on what is driving the red colors observed throughout the extent of these objects. A representative best-fit spectrum from each region is shown in the lower panel of Figure~\ref{fig:resolved_sps}. To generate these region-specific spectra, we first generate a rest-frame spectrum in FSPS for each UFO/region taken from the maximum of the parameter posteriors for that objects fit. Then, for each region, we stack the individual rest-frame spectra by taking a median at each wavelength/flux value to generate a ``median" spectrum for each region. It is clear that the two spectra are most similar at long wavelengths and least similar at shorter/bluer wavelengths with more blue flux in the outer regions consistent with the slight color gradients observed in UFOs that we showed in Figure~\ref{fig:color_grads}.

Turning to what drives these slight color gradients, in the upper panels of Figure~\ref{fig:resolved_sps}, we show the comparison between the inner and outer SED fitting results for the following parameters: stellar mass ($M_*$), visible attenuation from dust($A_V$), mass-weighted age (MWA), and specific star-formation rate (sSFR). All comparisons are shown as the ratio or difference of the inner parameter inference to the outer parameter inference as a function of the stellar mass inferred from the integrated fits (Section~\ref{sec:stellar_pops}). The red dashed line in each panel reflects the median ratio or difference between the inner and outer parameters, the gray dashed line is where the two regions have the same value, and the red shaded region shows the $\sim 1\sigma$ spread in the ratio/difference. This information is additionally summarized in Table~\ref{tab:res_stellar_pops}, where we show the median and percentiles of the inner and outer stellar population parameters.

The largest radial differences are observed between the inner and outer $A_V$, where it is clear that the majority of UFOs have higher central $A_V$ than outer $A_V$ although both regions have high $A_V$ (Figure~\ref{fig:mwa_av}). Taken at face value, this implies that dust concentrations are higher in the center as has been found before in studies of other galaxy populations \citep[e.g.,][]{nelson2016,whitaker2017,tacchella2018}. On the other hand, the gradients in the other parameters, MWA and sSFR, are not quite as strong, especially for the MWA which has a median ratio of $\sim$ one, but with significant scatter about the median (typical spread of $\sim$ 0.7). 
Here, the slight color gradients of our UFOs appear to be primarily caused by stronger central dust attenuation consistent with what was found in \citet{miller2023} for HST-selected star-forming galaxies.

\begin{center}
\input{inner_outer_table}
\end{center}

\section{UFO Structure} \label{sec:structure}
\subsection{Sizes and S\'ersic indices}
\begin{figure*}
    \includegraphics[width=0.24\textwidth]{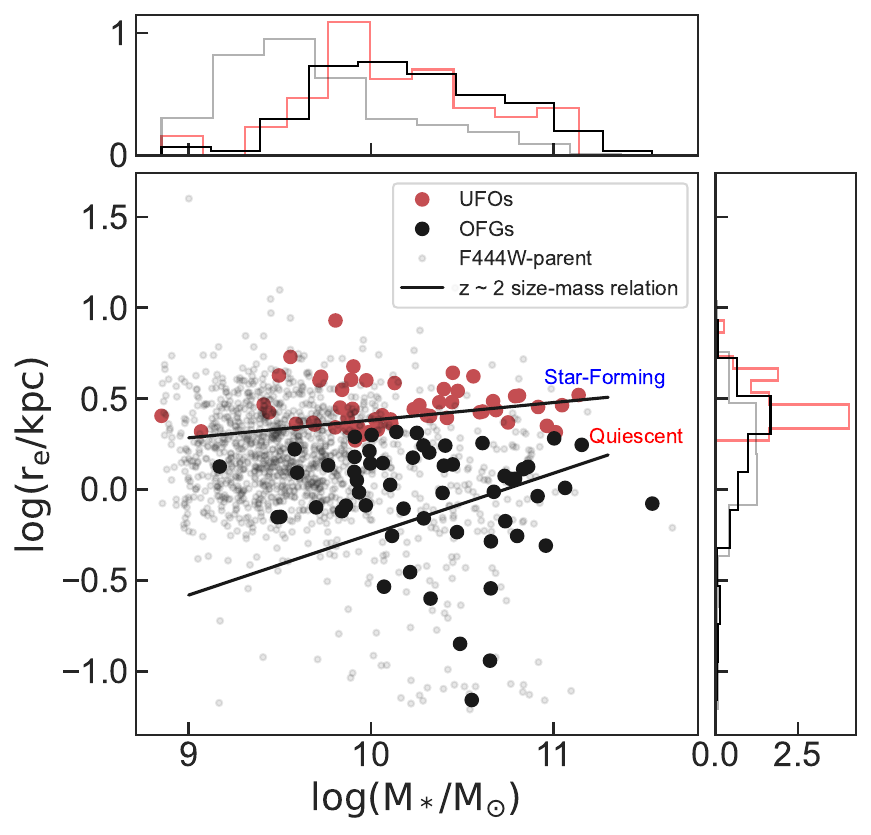}
    \includegraphics[width=0.23\textwidth]{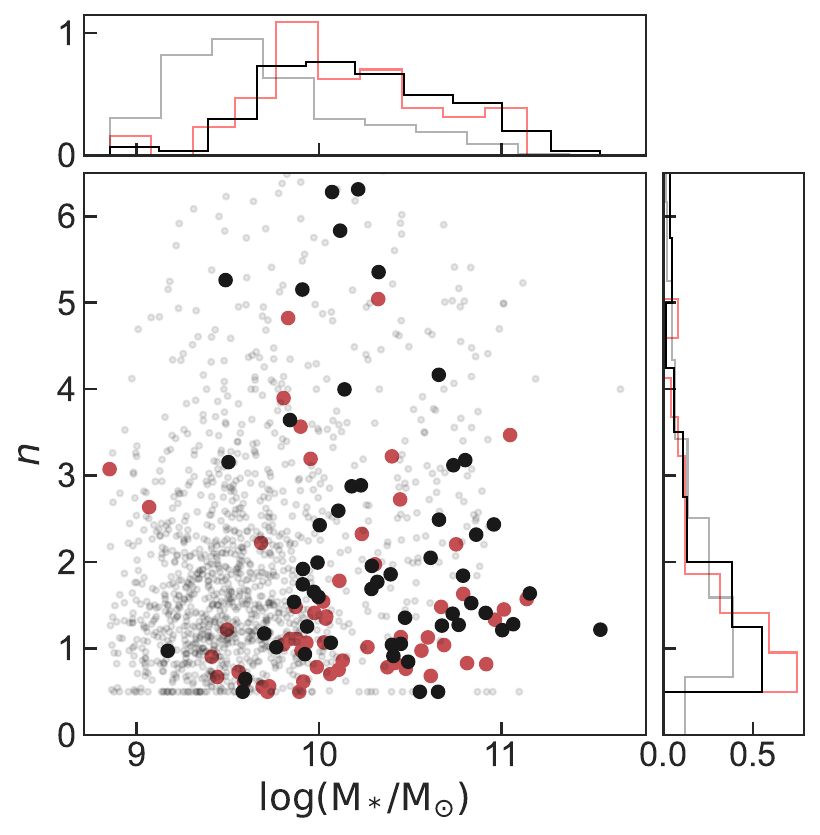}
    \includegraphics[width=0.23\textwidth]{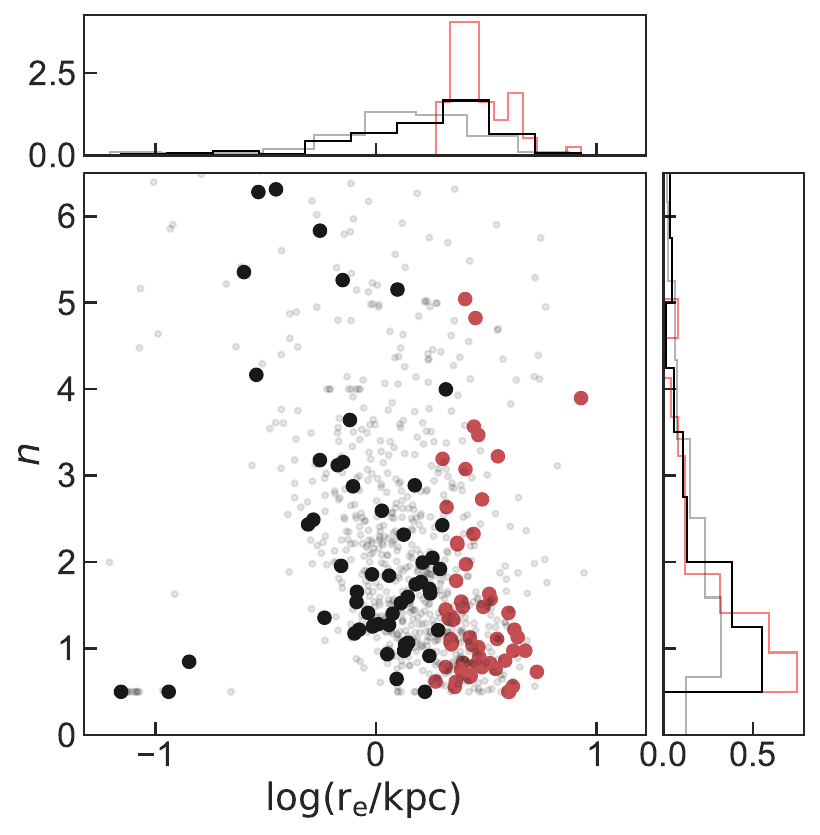}
    \includegraphics[width=0.23\textwidth]{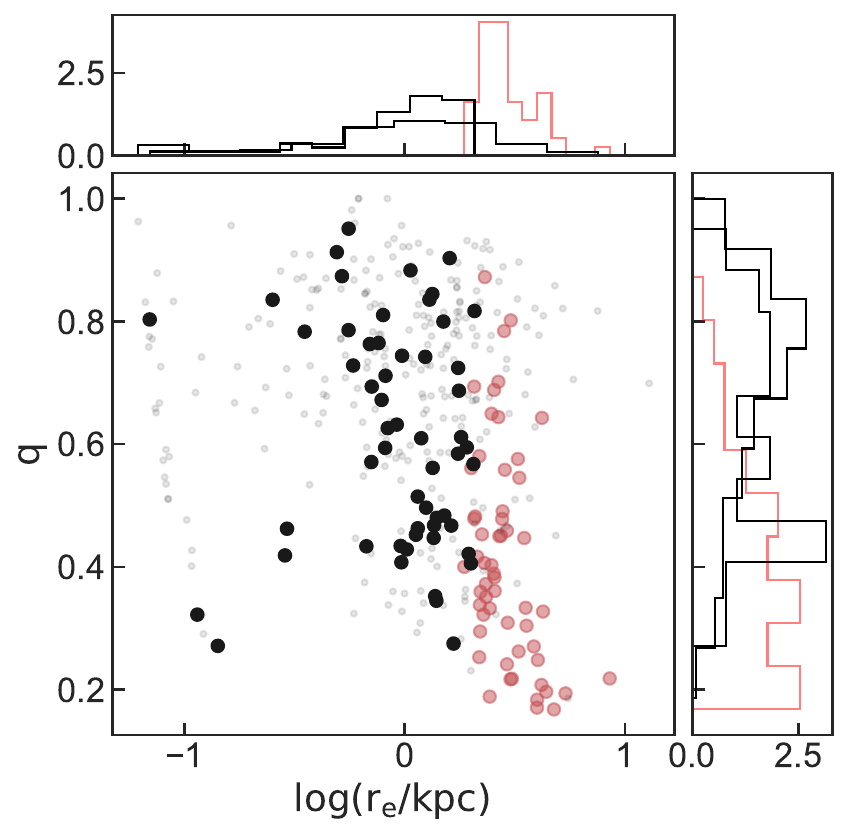}
    \caption{From left to right, we show how the UFOs, smaller optically-faint galaxies (OFGs), and bluer F444W-parent samples are distributed in the planes of size $-$ mass, S\'ersic index $-$ mass, S\'ersic index $-$ size, and axis ratio $-$ size. The histograms for the OFGs include the contribution from UFOs and the F444W-parent histograms include all galaxies in our samples. The solid black lines in the size-mass panel show the size-mass relationships from \citet{suess2019a} demonstrating that UFOs generally lie on or slightly above the star-forming relation while the OFGs are scattered around the quiescent size-mass relation (which is largely a result of our size cut). From the other two panels, we can see that the optically-faint galaxies have a much higher density of $n < 2$ objects, particularly for UFOs, relative to the bluer F444W-parent sample. There are no strong trends between the S\'ersic index and stellar mass or effective size. The right most panel shows $q$ versus $r_e$ illustrating that optically-faint galaxies have a range of sizes and tend to have lower $q$ values (especially for the UFOs). }
    \label{fig:size-mass}
\end{figure*}

Inferred morphological parameters are shown in Figure~\ref{fig:size-mass} and the median/percentiles of these parameters are shown in Table~\ref{tab:morph_params}. In the leftmost panel of Figure~\ref{fig:size-mass}, we show how the UFOs and their optically faint and F444W selected parent populations are situated in the size-mass plane. For comparison, we show the $z = 2$ size-mass relations from \citet{suess2019a} for star-forming galaxies and quiescent galaxies. Our sample of UFOs lie on or above the star-forming galaxy size-mass relation, showing that these are truly extended objects, both physically and apparently. Further, the full sample of optically faint galaxies is fairly evenly distributed in the size-mass plane relative to the parent sample. This is surprising. While we expected optically faint galaxies to have compact sizes to drive their high values of dust attenuation, this does not appear to be the case. The distribution of the sizes of the optically faint population is similar to that of the parent population. 

The middle two panels of Figure~\ref{fig:size-mass} show the distribution of S\'ersic indices versus stellar mass and effective size for the UFOs in the context of all optically faint galaxies and the F444W selected parent sample. The majority of the UFOs and the optically faint galaxy population in general have $n<2$, suggesting they have minimal structural contribution from a bulge. Contrary to our physical expectation that optically faint galaxies would be centrally concentrated to drive their high dust attenuations, the bulk of these objects have S\'ersic indices consistent with surface brightness profiles that are only slightly steeper than exponential. Interestingly, amongst F444W selected massive galaxies (${\rm log}(M_*/M_\odot) > 10$) with low S\'ersic indices ($n<2$), and $r_e>2$~kpc, 69\% are optically faint UFOs. Thus, neither compact sizes nor high S\'ersic indices can explain the high values of dust attenuation in these galaxies.

\begin{center}
\input{morph_param_table}
\end{center}

\subsection{3D shapes from axis ratio distribution modelling} \label{sec:ar_modeling}

\begin{figure}
    \includegraphics[width=0.5\textwidth]{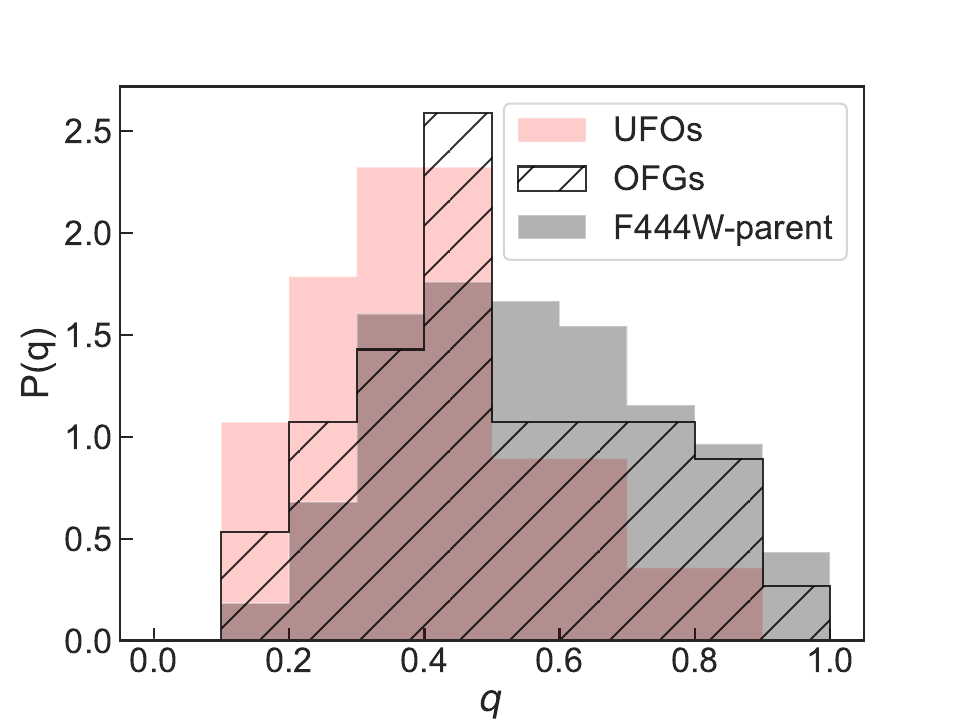}
    \caption{Axis-ratio distributions for the full F444W-parent sample, the optically-faint sample, and the subset of UFOs. Compared to the other distributions, the UFO distribution is much more skewed toward lower values of $q$ with the bulk of UFOs having $q < 0.4$. The other OFGs tend to have a fairly large number of low $q$ objects as well, but also have more $q > 0.5$ objects similar to what is seen in the full-parent sample.  }
    \label{fig:ar_dists}
\end{figure}

\begin{figure*}
    \includegraphics[width=1.0\textwidth]{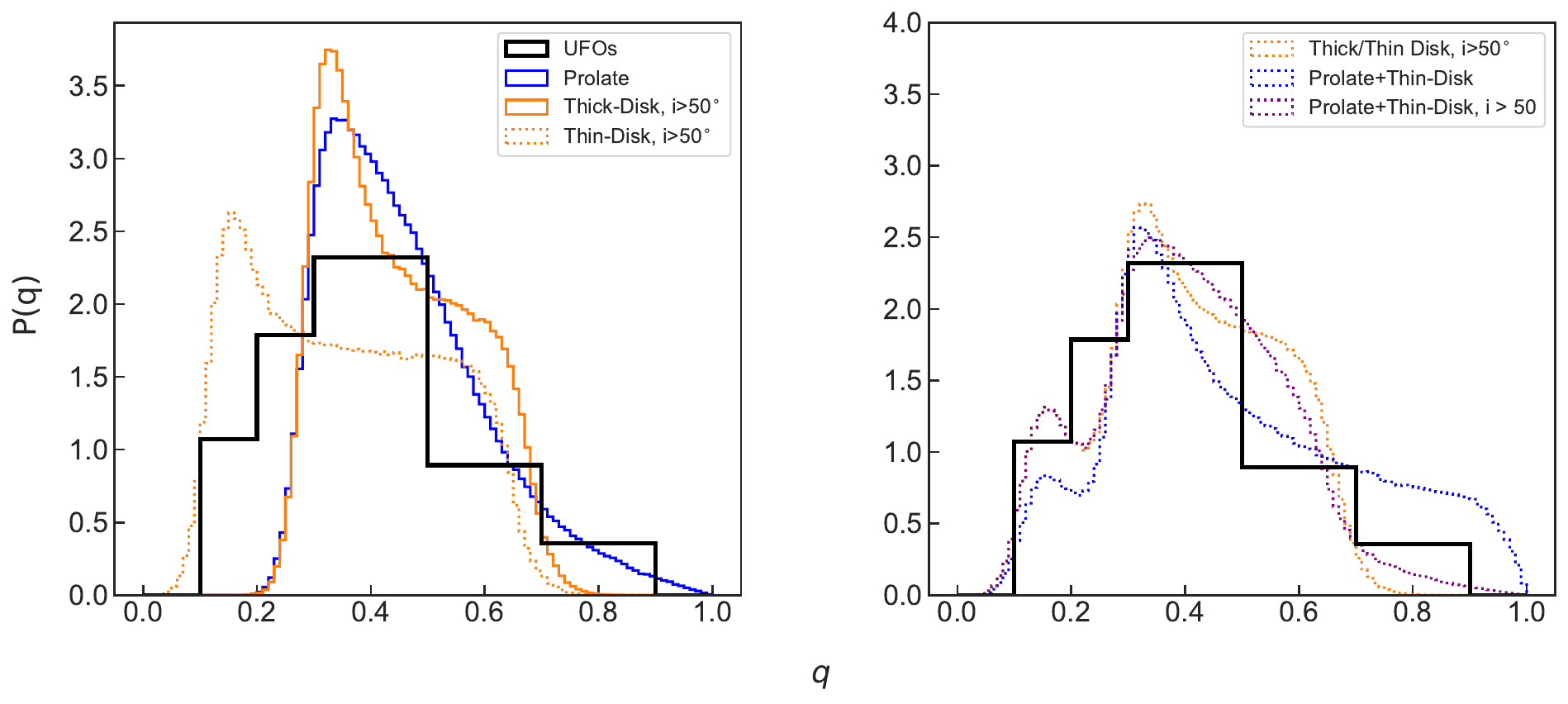}
    \caption{Left panel: Observed UFO axis-ratio distribution (black) compared with different $q$ distributions generated for  different mock galaxy populations (prolate/blue, thick-disk/solid orange, thin-disk/dashed orange) showing the plausible galaxy types that could give rise to the observed $q_{UFO}$ distribution. Right panel: Same as the left panel, but for different linear combinations of distinct populations (inclined thick/thin disks in orange, prolate/thin-disk in blue, and prolate/inclined thin disk in purple). It is difficult to reproduce $q_{UFO}$ without both an inclined thin-disk population and a prolate population. }
    \label{fig:mock_qs}
\end{figure*}

Here we explore the intrinsic shapes of the UFOs by modeling the distribution of their projected axis ratios \citep[e.g.][]{chang2013,vanderwel2014}.  In general, the shape of a galaxy can be approximated as an ellipsoid characterized by three axis lengths, $A \gtrsim B \gtrsim C$, which can be used to define the ellipticity (E) and triaxiality (T) parameters. Within the family of ellipsoids, there are three broad categories: the oblate ellipsoid ($A \sim B > C$, disky), prolate ellipsoid ($A > B \sim C$, elongated), or the spheroid ($A \sim B \sim C$). The projected axis-ratios of these shapes can be inferred from how how each ellipsoid type would appear in projection from a set of randomly drawn viewing angles with distinct shapes having unique projected axis-ratio distributions \citep[e.g.][]{chang2013,vanderwel2014} characterized by values of E and T and thus values of A, B, and C. For example, a $q$ distribution peaking between 0.8 and 1.0 would be predominately comprised of spheroids as they can only have large observed axis-ratios. A disk population would have a roughly flat $q$ distribution reflecting that disks can be observed edge-on, face-on, and everything in-between, while only observing inclined disks would result in an axis-ratio distribution without any higher $q$ objects. Prolate populations have axis-ratio distributions that peak at shorter values ($q \sim 0.4$) with a drop at $q \sim 0.2$ and a tail toward higher axis-ratios (see left panel of Figure~\ref{fig:mock_qs}).

In Figure~\ref{fig:ar_dists}, we plot the observed  axis-ratio distributions of the full F444W-selected parent sample, all optically faint galaxies, and the size-selected subset (the UFOs). We find that the UFOs typically have $q < 0.5$ and the smaller OFGs typically have $q \geq 0.5$. Taken at face value, this suggests that the UFOs are largely prolate or disk shaped and the more compact OFGs are more spheroid-dominated. We additionally perform a KS test on the UFO and OFG axis-ratio distributions to constrain the probability that these samples were drawn from the same underlying population. We find a p-value of $\sim 10^{-6}$ showing that, at the very least, the UFOs and OFGs represent different populations in terms of their axis-ratios, though we note that owing to their sizes being closer to the scale of the PSF, the shapes of the compact OFGs are much less certain. Figure~\ref{fig:mock_qs} shows the observed UFO axis ratio distribution compared to various mock populations. Our sample of UFOs has many more objects with $q \sim 0.1-0.2$ than the sample of \citet{nelson2023} and as such neither inclined thick-disk or prolate populations by themselves adequately match the observed UFO $q$ distribution. The tail of very low axis-ratios can only be achieved by having a population of intrinsically inclined thin-disks capable of reaching such low $q$ values. However, as can be seen in the right panel of Figure~\ref{fig:mock_qs}, an inclined thin-disk population by itself over-predicts these low $q$ objects and under-predicts those with higher $q$. For this reason, we show combinations of prolate and disk-like populations on the right. With the exception of the prolate+thin-disk (no inclination restriction) which overpredicts higher $q$ objects, the various combinations all seem to fairly well match the observed distribution making it challenging to uniquely assign an intrinsic shape to the UFOs. 

For these reasons, we turn to a modeling scheme that directly fits the observed axis-ratio distributions given some assumptions about the true underlying distribution. We use the BEAST (Price et al. in prep.) axis-ratio modeling code to perform nested sampling with \textsc{dynesty} \citep{speagle2020} to infer the intrinsic shapes of the UFOs. Here, we briefly summarize the functionality of this software. The ellipticity ($E = 1 – C^2$) and triaxiality ($T = \left[ 1 – B^2 \right]/\left[ 1 – C^2 \right]$) can be used to define specific sub-types of ellipsoids characterized by ranges of values of E and T reflecting the underlying relationships between A, B, and C. In practice, one can combine an observed $q$ distribution and, assuming the galaxies come from a population that can be described by a distribution of E and T, perform MCMC or nested sampling of the posterior to determine the most likely values for the parameters of the E and T distributions, and thus on the intrinsic shapes (i.e., oblate/disky or prolate) given these assumptions. 

\begin{figure*}
    \hspace*{-4ex}\includegraphics[width=0.33\textwidth]{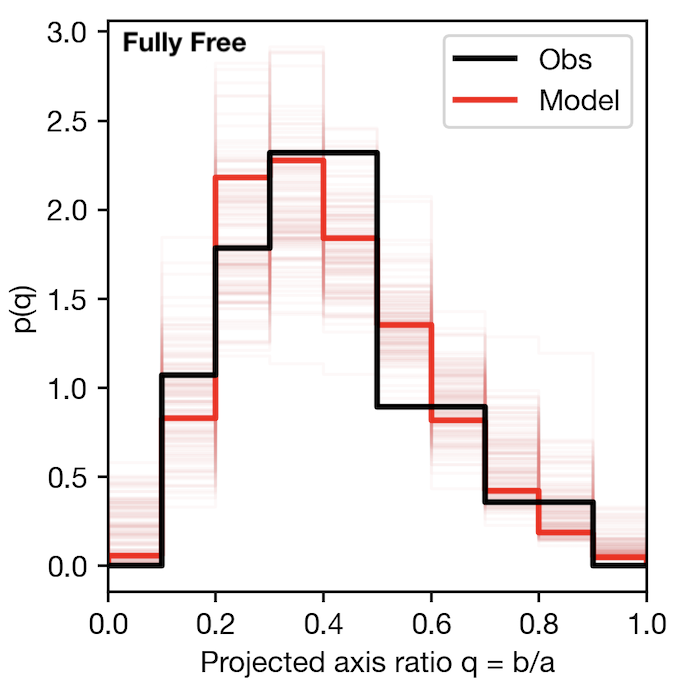}
    \hspace*{-4ex}\includegraphics[width=0.33\textwidth]{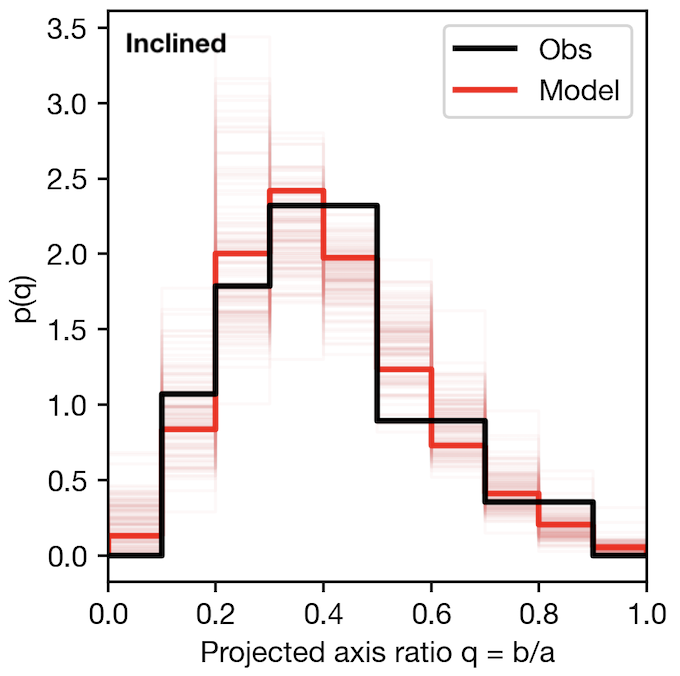}
    \hspace*{-4ex}\includegraphics[width=0.33\textwidth]{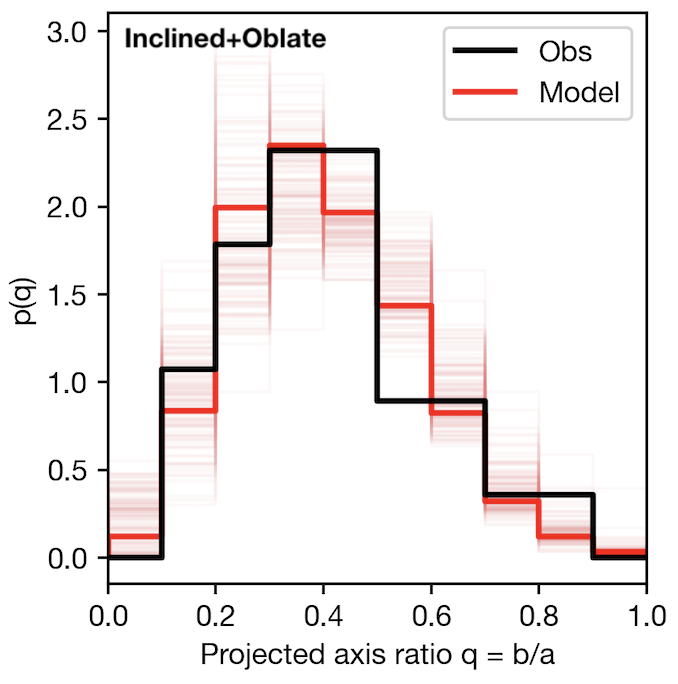}
    \hspace*{-4ex}\includegraphics[width=0.345\textwidth]{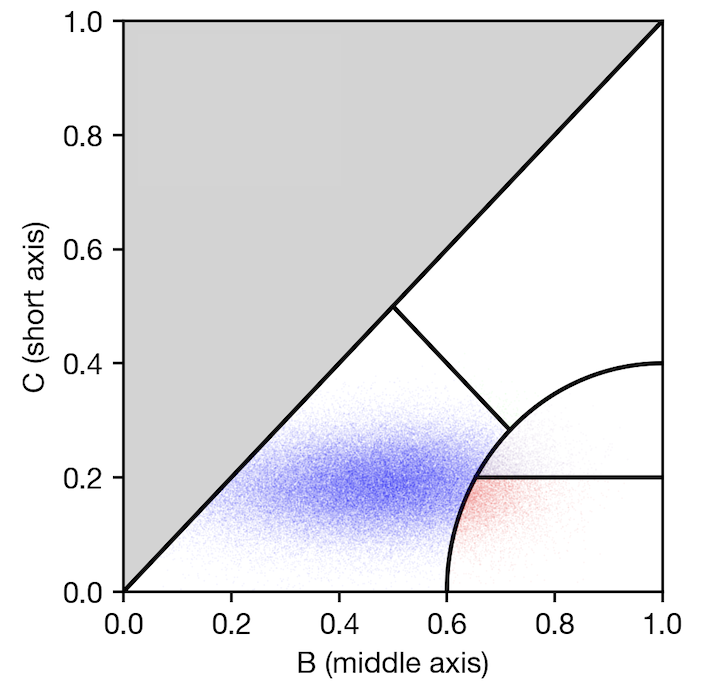}
    \hspace*{-4ex}\includegraphics[width=0.33\textwidth]{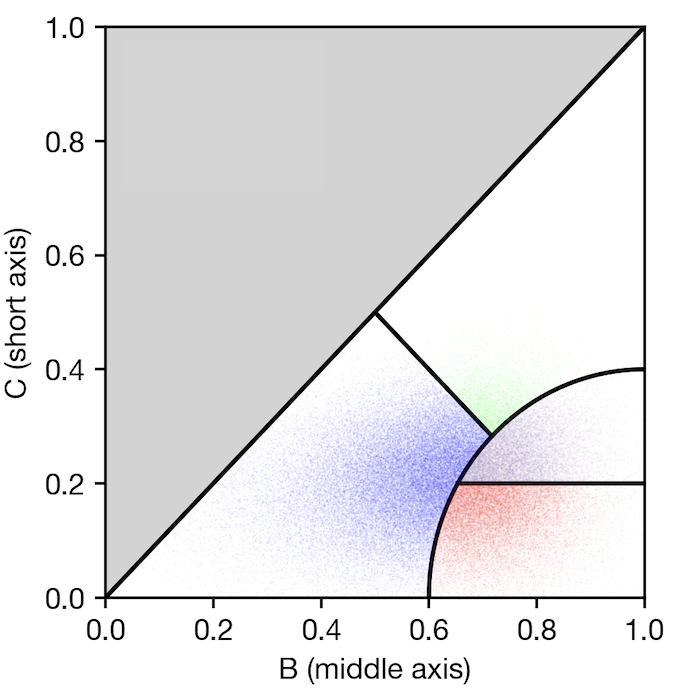}
    \hspace*{-4ex}\includegraphics[width=0.33\textwidth]{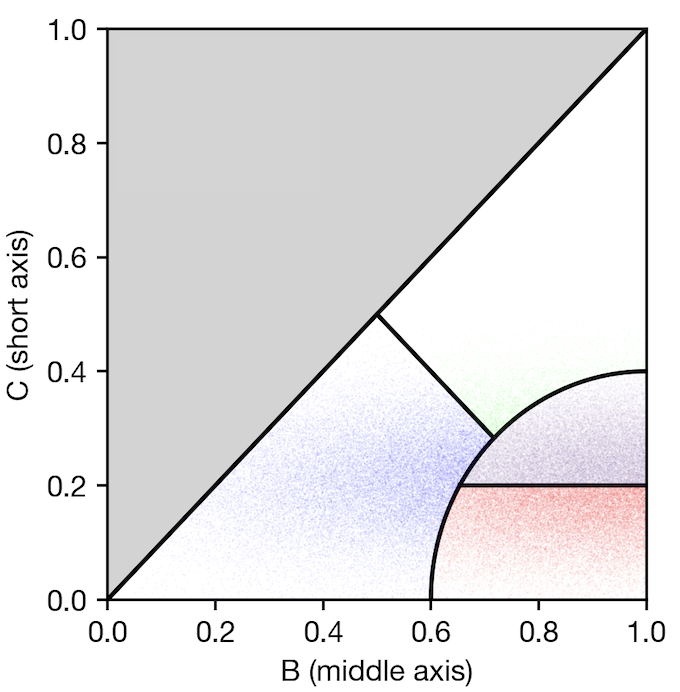}
    \includegraphics[width=0.30\textwidth]{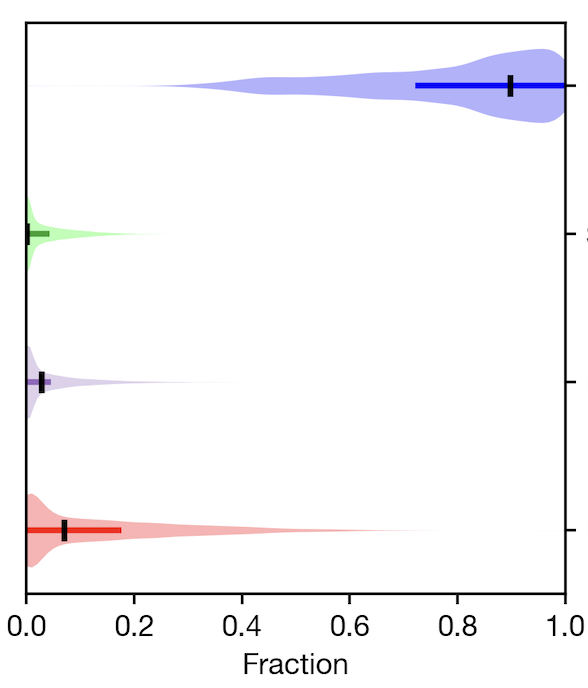}
    \hspace*{4ex}\includegraphics[width=0.30\textwidth]{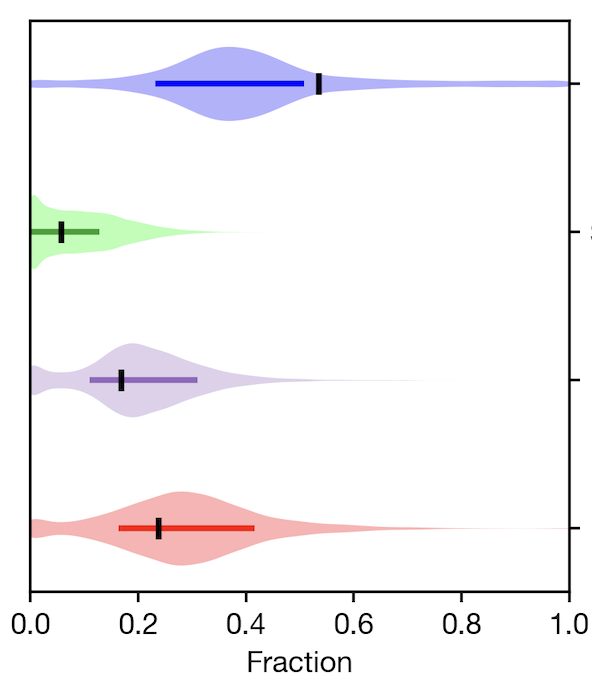}
    \hspace*{4ex}\includegraphics[width=0.38\textwidth]{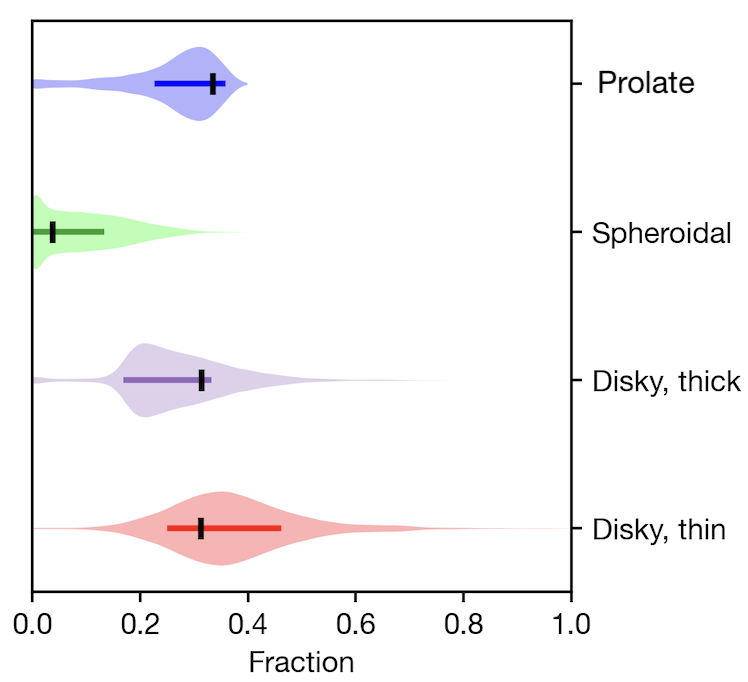}
    \caption{Summary of the results from our axis-ratio modeling with the BEAST. The top row shows the observed axis-ratio distributions in black compared with the best-fitting (maximum a posteriori) model distributions in red. Lighter red lines show $q$ distributions for random draws of the posterior distributions of $\mu_E, \mu_T, \sigma_E, \sigma_T$. The middle row shows random draws from the best-fitting model distribution in intrinsic axis-ratio space illustrating how the four distinct regions are defined and what the contribution from each region is. In the bottom row, we show the predicted fractions of four different galaxy shapes: elongated/prolate (blue), spheroidal (green), thick disk (purple) and thin disk (red). From left to right, we show the fully free version of the axis-ratio model, the model with restricted inclination, and the model with restricted inclination and a stronger preference for oblate galaxies. Here, vertical lines are the best-fit value for the fraction, horizontal lines show the 68\% confidence interval, and the violin plots (shaded areas) give a representation of the distribution with wider regions having more probability. }
    \label{fig:ar_modeling}
\end{figure*}

The BEAST modeling assumes all galaxies in the sample are drawn from a single population with Gaussian distributions of E and T, characterized by four free parameters: centers $\mu_E$ and $\mu_T$ and sigmas, $\sigma_E$ and $\sigma_T$. In our default model, we consider galaxies that are randomly observed from any viewing angle. However, we also employ a secondary model that restricts the inclination to be between $50-90^{\circ}$, and a third that additionally enforces oblateness (i.e., diskiness) by restricting the triaxiality parameter $< 0.33$, in addition to restricting the inclination. These secondary models encode the physical expectation that these objects are inclined disks. Each model fits for the $q$ distribution in 10 bins. For each parameter, we assume uniform priors from zero to unity except for the third model where we restrict the range of T to be between zero and 0.33. In what follows, we define four unique regions of intrinsic axis-ratio space (C versus B) that represent distinct shapes and can be seen in the middle panel of Figure~\ref{fig:ar_modeling}. These are spheroidal (green/upper right), prolate (blue/lower left), thin-disk (red/bottom of circular region), and thick-disk (purple/top of circular region). The results of fitting these three models are summarized in Figure~\ref{fig:ar_modeling} where we show the best-fit (maximum a posteriori) model $q$ distribution, random draws from the distribution of the best-fit model in intrinsic axis-ratio space, and the fractions of each shape that can be attributed to that single population model. 

From the top panels in Figure~\ref{fig:ar_modeling}, we can see that each different version of the model (free, inclined, inclined and oblate) does a reasonably good job at reproducing the observed UFO axis-ratio distribution. However, the model with no triaxiality or inclination constraints achieves this agreement in a markedly different way than the constrained models. In particular, when there are no constraints on the inclination or triaxiality of the population, the model prefers a population that is dominated by intrinsically prolate objects ($\sim90\%$) with only a $\sim 10\%$ contribution from disks (mostly thin) and essentially zero spheroidal objects. This tells us that the UFO $q$ distribution can be generated from a population that is dominated by prolate ellipsoids with a small contribution from thin-disks to reproduce the lowest $q$ UFOs. 

In the other two models, the story is very different. These disk assumption models can also produce axis-ratio distributions that are fairly consistent with the data showing that disk-based distributions are not inconsistent with the observations. In the middle panel of the bottom row, we can see that the restricted inclination model still predicts that prolate objects are the dominant contributor to the observed $q$ distribution, but only slightly with a $\sim 40\%$ contribution from thick and thin disks. Thus, even if we assume that we are only observing galaxies closer to edge-on, the axis-ratio modeling still statistically prefers a contribution from prolate objects. Moving to the third panel of the bottom row, we can see that if we restrict the triaxiality to be less than 0.33 as well as restricting the inclination, then the model predicts the dominant contribution coming from thick and thin disks ($\sim 60 \%$) with a contribution from prolate ellipsoids of at most $\sim 40\%$. This shows that even in the most disk-friendly modeling scenario, there is evidence for a contribution from prolate set of objects. That some optically faint objects might be prolate is a surprising finding as it is not straightforward how a prolate object viewed along its long-axis could have dust column densities large enough to render the galaxy optically faint. We can restrict the triaxiality even more, which will put more objects into the disk category, but this leads to much worse agreement between observed and model $q$ distributions and requires a $\sim 20\%$ spheroid contribution.

We note here a few modeling caveats that should be mentioned before we continue to discuss broader implications of these results. First, we assume that the intrinsic shapes of the UFOs can be accurately modeled as ellipsoids, which is a good approximation, but will clearly break down for certain objects. For example, ongoing galaxy mergers, galaxies with clumpy features, or galaxies with spiral arms and/or dust lanes could deviate somewhat from the ellipsoidal shape assumption. In fact, in some of our UFOs (Figure~\ref{fig:gallery}), we see evidence of clumps (29312), spiral features (187160), or multiple components / possible mergers (e.g., 167032 and 172813). These non-smooth components could impact the ensemble axis-ratio distribution leading to higher or lower inferences on the fractions of one or more shape subcategories. Additionally, if the UFOs are not a unified population of intrinsic shapes that can be characterized by a single distribution, then our modeling assumptions start to break down and the inferred parameters are not necessarily telling us about the true breakdown of intrinsic shapes. 

\section{Discussion} \label{sec:disc}

This paper presents an analysis of the stellar populations and structural properties of an unexpected population of elongated, optically faint galaxies from the JADES survey. Here, we discuss why these results are surprising in terms of existing studies of the intrinsic shapes of galaxies, what drives the optical faintness of these objects, and their subsequent evolution.

Probably the most surprising aspect of this population of objects is their structure: they have fairly large radii, nearly exponential light distributions, and are highly elongated. Beginning with their sizes, the leftmost panel of Figure~\ref{fig:size-mass} shows that a population of optically faint galaxies exist (which we call UFOs) whose sizes are larger than average star-forming galaxies at their masses and redshifts. This is surprising because at a fixed dust mass, galaxies with a dust distribution that is more concentrated will have higher dust column densities and hence larger values of dust attenuation \citep[e.g.][]{nelson2014}. These higher dust column densities could come from compact sizes or concentrated distributions (i.e., high S\'ersic indices). As such, we may expect optically faint galaxies to preferentially have small sizes and / or high S\'ersic indices. Thus, while the naive physical expectation for the structures of optically faint galaxies would be to have compact sizes to drive their high values of dust attenuation, this does not appear to be the case with the distributions of the optically-faint and bluer F444W selected parent samples being similar. Specifically, amongst objects with stellar masses of greater than $10^{10}$\msun, sizes $>2$kpc, 39\% are optically faint while amongst objects with sizes less than 2kpc, 22\% are optically faint. Optical faintness does not appear to correlate with size. This is surprising given the findings of \citet{gomezguijarro2023} who find that amongst star forming galaxies at $3<z<7.5$, those with $A_V>1$ have $\sim30\%$ smaller sizes. Although, recent simulations find that orientation (i.e., viewing angle) and not physical characteristics, is the main determinate of whether a galaxy will be optically faint or not, with galaxies only appearing optically-faint in some fraction (depending on the object) of the orientations \citep{cochrane2023}.

The second surprise is their S\'ersic indices: the bulk of the UFOs and the optically faint galaxies in general have $n<2$ (see Fig.~\ref{fig:size-mass}). Previous multi-wavelength studies looking at the morphologies of dusty galaxies in both optical/UV and sub-mm wavelengths general find that these sources are more compact at longer observed wavelengths \citep[e.g.,][]{hodge2016,tadaki2017,nelson2019} suggesting that these galaxies may be in the process of building dense stellar bulges \citep[e.g.,][]{lang2014,whitaker2017} with dust attenuation causing these bulges to be missed at shorter wavelengths. Bulge-dominated galaxies are typically characterized by high S\'ersic indices reflecting more centrally concentrated light profiles, which is the opposite of what we see in the sample of optically faint galaxies studied in this paper. The UFOs and optically-faint galaxies in our sample have light distributions more indicative of exponential profiles suggesting they are strongly disk dominated (though prolate is also a possibility). The massive portion of the galaxies in our sample are approaching stellar masses at which they will likely quench yet most show little evidence for having or building bulges which appears to be a prerequisite for quenching \citep[e.g.][]{lang2014,whitaker2017}. Although most of the UFOs have low S\'ersic indices in F444W emission, it is possible that even the F444W light is attenuated in the central regions and a bulge component could appear in even redder wavelengths. MIRI or ALMA data would be needed in order to definitively rule out the presence of a highly obscured bulge component.  

The third surprise is that most of the UFOs are elongated. These objects are selected based on their fluxes, colors, and sizes; there is no explicit selection for them to be elongated. The most obvious explanation perhaps for a highly reddened cohort of large galaxies to be elongated is that we are looking at an edge-on subset of a disk galaxy population. Another possible explanation for high dust attenuation is geometric: if galaxies are intrinsically disky, then the dust column density will be much larger when viewed edge-on than face-on \citep[e.g.][]{wild2011,patel2012}. If the UFOs are in fact disk-dominated objects, viewing them edge-on could explain why they are optically faint and would not have appeared in previous studies.  

While there is significant physical reason to believe the UFOs are intrinsically edge-on disks, axis-ratio modelling suggests a significant fraction of objects may be intrinsically \textit{prolate} instead of oblate/disky. This type of statistical modeling was not possible in \citet{nelson2023} given the small sample size; having the large sample provided by JADES is essential. In a framework assuming UFOs are a single population, axis-ratio modelling suggests that at least 30\% of UFOs are intrinsically prolate. This is very surprising.  In an intrinsically prolate population, a physical argument akin to the higher dust columns in inclined disk galaxies, would suggest a preference for optical faintness when looking at the prolate object with the long axis along the line of sight. In this configuration, the object would appear small and circular (perhaps like some of the smaller optically-faint galaxies in our sample) and it would physically make sense for these objects to have large dust columns. Viewed with the long axis perpendicular to the line of sight, it is harder to understand how dust attenuation could be strong enough to produce optical faintness. 
 
 \begin{figure}
    \includegraphics[width=0.5\textwidth]{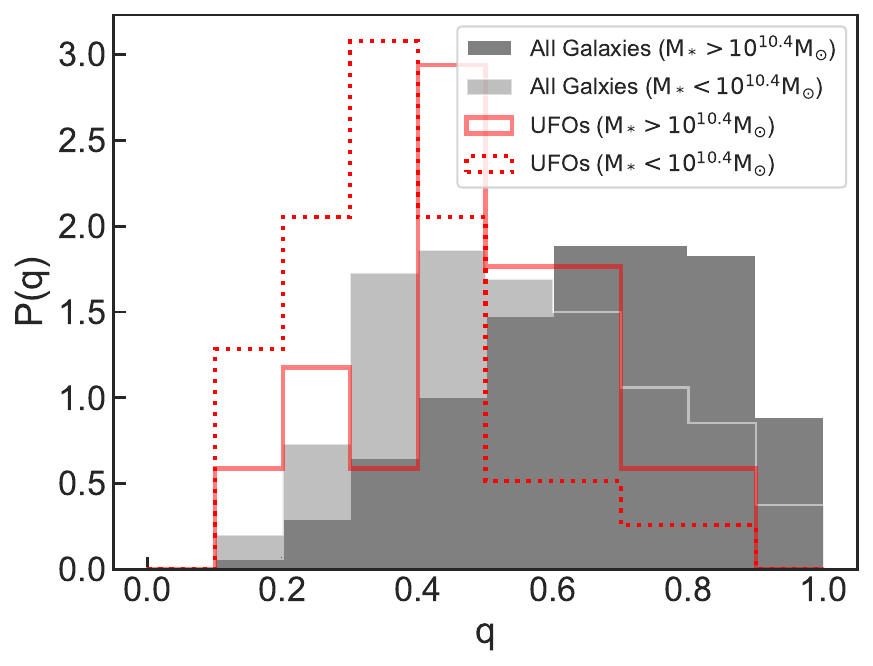}
    \caption{Axis-ratio distributions for the UFOs and the entire sample of galaxies considered in this work split into low mass ($\mathrm{M_* < 10^{10.4} M_{\odot}}$) and high mass ($\mathrm{M_* < 10^{10.4} M_{\odot}}$) samples. It can clearly be seen that the majority of the low $q$ objects in our whole sample are predominately coming from lower mass UFOs, but with a strong contribution from the higher mass UFOs as well. The $q$ distribution of low-mass UFOs is largely consistent with a prolate population, but the lower sample size of the high-mass group make it harder to come to strong conclusions about the intrinsic shapes of the higher mass UFOs. }
    \label{fig:diffm_qdists}
\end{figure}

To put these results in context, we compare to previous studies of the intrinsic shapes of galaxies as a function of redshift and stellar mass. Our stellar population modelling suggests that these objects have fairly high stellar masses; observations of star-forming galaxies (SFGs) with HST over the redshift range $1 < z < 3$ have shown that high mass ($M_* \ge 10^{10} M_{\odot}$) SFGs are predominately disks \citep[e.g.,][]{vanderwel2014,zhang2019} especially at $z<2$. However, at higher redshifts ($z \gtrsim 2$) and lower masses ($M_* \le 10^{9.5} M_{\odot}$) the shapes of SFGs become dominated by prolate objects. In other words, the fraction of star-forming prolate galaxies increases towards higher redshifts and lower masses, while the opposite is true of star-forming disks.

For this reason, and given the fairly wide range of inferred UFO stellar masses, we consider the possibility that the UFOs are composed of a lower mass population and a higher mass population with distinct axis-ratio distributions. Figure~\ref{fig:diffm_qdists} shows the axis ratio distributions of UFOs and the parent population of F444W selected galaxies divided by mass at $\mathrm{M_* = 10^{10.4} M_{\odot}}$. As expected, the axis-ratio distribution of the low mass UFOs is more skewed towards lower $q$ values than that of the high mass UFOs and axis-ratio modeling suggests nearly 100\% are prolate. There are not enough galaxies in the high mass bin to determine their intrinsic shapes from statistical modelling but their axis-ratio distribution is certainly not that of a purely prolate population.

It has been shown in \citet{vegaferrero2023}, based on a machine-learning based classification scheme for inferring galaxy morphologies, that approximately half of galaxies visual classified as disks from CEERS imaging are more consistent with prolate or spheroidal populations than being true disks. Thus, we may indeed be looking at some objects that visually appear to be disks, but in reality are prolate. Further, it is worth noting that populations of massive, but quenched prolate galaxies at $z\sim0$ have been identified in cosmological simulations \citep[e.g., ][]{ebrova2017,thob2019} as well as observationally \citep[e.g.,][]{tsatsi2017}. \citet{ebrova2017} suggests that these prolate objects are either formed prolate (or become prolate early in their lifetimes) or formed through radially aligned mergers that induce the stellar shapes towards prolate ellipsoids. These types of observations could be consistent with our plausible finding of star-forming prolate intermediate mass ($\mathrm{M_* < 10^{10.4} M_{\odot}}$ galaxies at $2<z<6$ which could then be the progenitors of massive quenched galaxies at $z\sim0$. Although we can rule out that the UFOs are not a bulge-dominated population in F444W light, definitively determining the intrinsic shapes of these objects will require kinematic measurements.

Turning back to why these objects are optically-faint, in Figure \ref{fig:mwa_av} we show the distribution of optically faint galaxies and our F444W-selected parent sample in a plane of two likely culprits: age and dust attenuation, which both act to redden the SEDs of galaxies. The most significant effect is dust attenuation: most galaxies at $2<z<6$ and $F444W < 24.5$ that have $A_v > 2$ are optically faint. There is a secondary trend with age in which the optically faint galaxies selected from JADES are on average slightly older than the parent population but this is decidedly subdominant to the dust attenuation. Further, as shown in Fig.~\ref{fig:size-mass}, 58\% of galaxies in our F444W-selected parent sample with $q<0.3$ are optically faint. On the other hand, both size and S\'ersic index do not appear driving optical faintness (at least as defined in this paper). Thus, having a low projected axis-ratio appears to be the best predictor of optical faintness amongst the standard morphological indicators $-$  effective radius, S\'ersic index, and axis-ratio. That said, a very small fraction of the compact optically-faint galaxies (i.e. non-UFOs) have $q<0.3$ so inclination does not appear to provide an explanation for the optical faintness of these more compact objects. 

Finally, we consider how this enigmatic galaxy population may evolve toward lower redshifts. With stellar masses of \logm $>10$ at $z\sim2-3$, UFOs are likely to evolve into quiescent galaxies with \logm$>11$ in the local universe \citep[e.g.][]{behroozi2019}. In fact, in Figure~\ref{fig:stellar_pop_comp}, we can see that a small fraction of our three samples have SFRs up to $\sim$ 3 dex below the $z = 2.5$ SFMS from \citet{leja2022} suggesting that these galaxies have already ceased their star-formation. This is a sharp transition to quiescence at a characteristic mass of \logm $\sim$ 10.4 consistent with previous studies finding that quenching is efficient at a similar mass \citep[e.g.,][]{contini2020}. With the typically small quantity of dust attenuation in quiescent galaxies in the local universe, the dust attenuation seen in these objects at $z>2$ will need to almost completely disappear in the intervening time \citep[e.g.][]{whitaker2021}. Significant structural change is also likely as they need to transition from prolate or oblate morphologies to having significant to dominant bulge components. 

\section{Conclusions} \label{sec:concl}

We have performed a morphological and stellar population analysis of a sample of 112 optically-faint galaxies with a specific focus on the elongated sub-set of optically faint galaxies, dubbed UFOs. Specifically, we explored how these galaxies are situated in certain fundamental galaxy scaling relations such as the size-mass plane and the star-forming main-sequence as well as investigations into how the stellar populations vary between the inner and outer regions and finally an in depth look at the morphological parameters of these objects such as their sizes, S\'ersic indices, and axis-ratios. Our main goal has been to place stronger constraints on the nature of the UFO galaxy population and to place the optically faint and UFO populations within the larger context of the massive, star-forming galaxy population. Our main results are summarized below.

\begin{enumerate}

\item We identify  112 optically-faint objects in the JADES survey of which 56 are classified as UFOs with apparent sizes larger than 0.25". The UFOs are typically between $2 < z < 4$ with a range of stellar masses ($M_* \sim 10^{9-11} M_{\odot}$), high SFRs, and high $A_V$. Compared to a mass and redshift matched sample of bluer F444W-selected galaxies in JADES, the UFOs tend to have higher stellar masses and SFRs, but have much lower stellar masses than the Herschel detected ultrared galaxies from \citet{ma2019}.

\item UFOs have red colors throughout the extent of their bodies, but still possess slight negative color gradients with outskirts that are less red than the interiors. These color gradients are likely driven by increased central dust concentration and not by the ages of the stellar populations.

\item The observed UFO axis-ratio distribution in F444W is consistent with observing some combination of randomly oriented disks and randomly oriented prolate galaxies. We perform a detailed Bayesian modeling of the $q$ distribution using the BEAST and find that the UFOs can not be uniquely said to consist of one single galaxy shape. It is possible that the UFOs consist of a lower mass prolate population and a higher mass oblate/disk population, but without a larger sample it is hard to verify this.

\item We find that the strongest predictors of optical faintness in the galaxy populations are  $A_V$ and $q$. That increasing $A_V$ is a strong predictor of optical faintness is not surprising given that larger dust attenuation will naturally lead to less optical light escaping a galaxy. Morphologically, it is somewhat surprising that the strongest predictor for optical faintness is a low $q$ as it was thought that small physical size would be the easiest way to drive large enough dust columns to make galaxies optically-faint, but that is not what we see given the large sizes of UFOs. 
\end{enumerate}

UFOs are an unanticipated galaxy population that JWST has made easy to reveal and study. The existence of optically-faint galaxies with extended structure was surprising owing to the expectation that only compact sizes could lead to large enough dust columns and as early JWST identified optically-faint galaxies were found to be relatively compact \citep{gomezguijarro2023}. The suggestion that a significant fraction of these objects are prolate instead of disks is an interesting possibility with implications for the $z > 2$ shapes of star-forming galaxies and how these shapes evolve to the present day. Future work will increase these samples to larger numbers and seek stronger constraints on the morphologies with kinematics from JWST spectra allowing us to come to a definitive conclusion on the intrinsic 3D shapes of UFOs. 

\bigspace 
J.L.G gratefully acknowledges support provided by NASA through grants 20-ASTRO20-0200 and HST-AR-16146 and support from the NRAO through grant SOSPADA-025. E.J.N acknowledges support provided by NASA through grants HST-AR-16146 and JWST-GO-01895. The research of C.C.W is supported by NOIRLab, which is managed by the Association of Universities for Research in Astronomy (AURA) under a cooperative agreement with the National Science Foundation. S.H.P acknowledges support through grant JWST-GO-2561. The Cosmic Dawn Center (DAWN) is funded by the Danish National Research Foundation under grant No. 140. A.J.B acknowledge funding from the FirstGalaxies Advanced Grant from the European Research Council (ERC) under the European Union’s Horizon 2020 research and innovation programme (Grant agreement No. 789056). W.B. acknowledges support by the Science and Technology Facilities Council (STFC), ERC Advanced Grant 695671 "QUENCH". K.B. acknowledges support from the Australian Research Council Centre of Excellence for All Sky Astrophysics in 3 Dimensions (ASTRO 3D), through project number CE170100013. E.C.L acknowledges support of an STFC Webb Fellowship (ST/W001438/1). D.J.E is supported as a Simons Investigator and by JWST/NIRCam contract to the University of Arizona, NAS5-02015. R.H acknowledges support from the Johns Hopkins University, Institute for Data Intensive Engineering and Science (IDIES). R.M. acknowledges support by the Science and Technology Facilities Council (STFC), by the ERC through Advanced Grant 695671 QUENCH, and by the UKRI Frontier Research grant RISEandFALL. R.M. also acknowledges funding from a research professorship from the Royal Society. B.D.J., G.R., M.R., and B.E.R. acknowledge support from a JWST/NIRCam contract to the University of Arizona NAS5-02015. This work utilized the Alpine high performance computing resource at the University of Colorado Boulder. Alpine is jointly funded by the University of Colorado Boulder, the University of Colorado Anschutz, and Colorado State University

\vspace{5mm}
\facilities{JWST(NIRCam)}


\software{astropy \citep{astropy,astropy2018,astropy2022}, \textsc{Dynesty} \citep{speagle2020}, \textsc{GALFIT} \citep{peng2002,peng2010}, \textsc{Lenstronomy} \citep{birrer2018}, matplotlib \citep{matplotlib}, numpy \citep{numpy}, photutils \citep{photutils}, \textsc{Prospector} \citep{johnson2021}
          }

\bibliography{paper_refs}{}
\bibliographystyle{aasjournal}

\end{document}

%% file: int_stellar_pop_props.tex
\startlongtable
\begin{deluxetable*}{lcccc}
\tabcolsep=5pt
\tablecolumns{5}
\tablewidth{0pc}
\tablecaption{Integrated Stellar Population Parameters \label{tab:int_stellar_pops}}
\tabletypesize{\footnotesize}
\tablehead{Parameter  & \colhead{UFOs}   & \colhead{OFGs} & \colhead{F444W-parent} & \colhead{Herschel Ultrared}    }
\startdata
z & $2.42^{+0.58}_{-0.50}$ & $2.78^{+0.85}_{-0.64}$ & $2.19^{+0.68}_{-0.49}$ & $3.24^{+1.10}_{-0.51}$ \\
$\mathrm{log(M_*/M_{\odot})}$ & $10.04^{+0.63}_{-0.33}$ & $10.29^{+0.50}_{-0.39}$ & $9.81^{+0.63}_{-0.21}$ & $11.57^{+0.18}_{-0.42}$ \\
$\mathrm{log(SFR/(M_{\odot}/yr))}$ & $1.58^{+0.50}_{-0.37}$ & $1.48^{+0.48}_{-0.57}$ & $0.94^{+0.51}_{-1.27}$ & $2.86^{+0.23}_{-0.47}$ \\
$A_V$ & $2.76^{+0.47}_{-0.43}$ & $2.61^{+0.52}_{-0.65}$ & $0.87^{+0.62}_{-0.40}$ & -- \\
MWA [Gyr] & $0.97^{+1.43}_{-0.51}$ & $0.90^{+1.50}_{-0.51}$ & $1.09^{+1.88}_{-0.35}$ & -- \\
\enddata
\tablecomments{
 Reported quantity is the median and percentiles (16th and 84th) of the distribution for that parameter.
 }

\end{deluxetable*}

%% file: inner_outer_table.tex
\startlongtable
\begin{deluxetable}{lcc}
\tabcolsep=8pt
\tablecolumns{4}
\tablewidth{0pc}
\tablecaption{Inner and Outer Stellar Population Parameters\label{tab:res_stellar_pops}}
\tabletypesize{\footnotesize}
\tablehead{Parameter  & \colhead{Inner}   & \colhead{Outer}     }
\startdata
$\mathrm{log(M_*/M_{\odot})}$ & $9.86^{+0.48}_{-0.42}$ & $10.11^{+0.42}_{-0.47}$ \\
$\mathrm{log(sSFR/yr^{-1})}$ & $-8.28^{+0.55}_{-0.70}$ & $-8.64^{+0.58}_{-0.78}$ \\
$A_V$ & $4.26^{+1.46}_{-0.70}$ & $2.29^{+0.32}_{-0.57}$ \\
MWA [Gyr] & $0.94^{+0.49}_{-0.54}$ & $1.08^{+0.44}_{-0.57}$ \\
\enddata
\tablecomments{
 Reported quantity is the median and percentiles (16th and 84th) of the distribution for that parameter.
 }

\end{deluxetable}

%% file: morph_param_table.tex
\startlongtable
\begin{deluxetable}{lcc}
\tabcolsep=8pt
\tablecolumns{4}
\tablewidth{0pc}
\tablecaption{Morphological Parameter Summary \label{tab:morph_params}}
\tabletypesize{\footnotesize}
\tablehead{Parameter  & \colhead{UFOs}   & \colhead{OFGs}     }
\startdata
$\mathrm{log(r_e/kpc)}$ & $0.43^{+0.17}_{-0.09}$ & $0.04^{+0.19}_{-0.30}$ \\
Axis-ratio (q) & $0.39^{+0.20}_{-0.17}$ & $0.60^{+0.21}_{-0.17}$ \\
S\'ersic Index (n) & $1.12^{+1.53}_{-0.39}$ & $1.72^{+1.99}_{-0.68}$ \\
\enddata
\tablecomments{
  Reported quantity is the median and percentiles (16th and 84th) of the distribution for that parameter.
 }

\end{deluxetable}